\documentclass[twocolumn]{aastex63}
\usepackage{fontawesome}
\usepackage{CJK}
\usepackage{natbib}
\usepackage{amsmath}
\usepackage{geometry}
\graphicspath{{figures}}

\newcommand{\teff}{T$_{\rm eff}$}
\newcommand{\feh}{\rm{[Fe/H]}}

\received{December 3, 2021}
\revised{---}
\accepted{---}
\submitjournal{AJ}

\shorttitle{Abundances For Exoplanet Characterization}
\shortauthors{Kolecki \& Wang}

\graphicspath{{./}{figures/}}

\begin{document}
%\nocite{*}
\begin{CJK*}{UTF8}{gbsn}
\newgeometry{lmargin = .75in, rmargin = .75in,bmargin = 1in,tmargin = 2in}

\title{Measuring Elemental Abundances of JWST Target Stars for Exoplanet Characterization \newline I. FGK Stars}

\author{Jared R. Kolecki}\affiliation{Department of Astronomy, The Ohio State University, Columbus, Ohio 43210, USA}

\author[0000-0002-4361-8885]{Ji Wang (王吉)}\affiliation{Department of Astronomy, The Ohio State University, Columbus, Ohio 43210, USA}

\correspondingauthor{Jared Kolecki}
\email{kolecki.4@osu.edu}

\begin{abstract}
With the launch of the JWST, we will obtain more precise data for exoplanets than ever before. However, this data can only inform and revolutionize our understanding of exoplanets when placed in the larger context of planet-star formation. Therefore, gaining a deeper understanding of their host stars is equally important and synergistic with the upcoming JWST data. We present detailed chemical abundance profiles of 17 FGK stars that will be observed in exoplanet-focused Cycle 1 JWST observer programs. The elements analyzed (C, N, O, Na, Mg, Si, S, K, and Fe) were specifically chosen as being informative to the composition and formation of planets. Using archival high-resolution spectra from a variety of sources, we perform an LTE equivalent width analysis to derive these abundances. We look to literature sources to correct the abundances for non-LTE effects, especially for O, S, and K, where the the corrections are large (often $> 0.2~\textrm{dex}$).
With these abundances and the ratios thereof, we will begin to paint clearer pictures of the planetary systems analyzed by this work. With our analysis, we can gain insight into the composition and extent of migration of Hot Jupiters, as well as the possibility of carbon-rich terrestrial worlds.

\end{abstract}
\section{Introduction}\label{Introduction}

Advances made in the last decade within the field of exoplanets have allowed us, for the first time,
to characterize exoplanet atmospheres via their spectra by transit, emission, and Doppler spectroscopy \citep[See a review by][]{Madhusudhan2019}. 

With the launch of the James Webb Space Telescope \citep[JWST,][]{JWST} in late 2021, as well as planned missions such as Twinkle \citep{Twinkle} in 2024, and ARIEL \citep{ARIEL} in 2028, all of which will have spectroscopic capabilities, it will be possible to probe exoplanet atmospheres in more detail than ever, allowing for insight into atomic and molecular composition, pressure/temperature profiles, and the presence of clouds/haze.

Properties of the atmosphere of a planet's host star have significant impact on the planet formation process (as the host star composition can be used as a proxy for the composition of the initial proto-planetary disk from which the planets were born). By comparing planetary and stellar atmospheric chemical composition, we can gain insight into where and how planets form and migrate \citep[e.g.][]{Oberg+2011,Madhusudhan+2017,Turrini+2021}.

In this paper, we have aimed to carry out abundance analyses for many of the most notable elemental species involved in planet formation (namely, C, N, O, Na, Mg, Si, S, K, and Fe.) These analyses were performed on stars selected from JWST Cycle 1 GTO and General Observer programs in the Extra-Solar Planets category. Our targets were chosen such that all had high fidelity archival spectra available. The abundances derived in this paper will be critical in comparing stellar and planetary atmospheric chemical abundances for probing planet formation. 

This paper focuses on 17 FGK stars set to be observed by JWST. A separate paper (Kolecki et al., in prep.) will consider a set of JWST M-dwarf targets, which have more complex spectra that require different analytical techniques.
 For many of the stars in our sample, no similar extensive, homogeneous abundance analysis has been done previously involving our selected elements. In the extreme cases (e.g. TOI-193) there have been no such analyses published at all. Seven of our stars are present in the Hypatia catalog \citep{Hinkel+2014}\footnote{available at \url{https://www.hypatiacatalog.com/}}, which provides a number of abundance measurements. 

However, that catalog is a compilation of several literature sources, which may not provide complete coverage of all the elements analyzed in this work. Furthermore, the differing methods among these sources can lead to issues when comparing, for example, abundances derived from LTE versus non-LTE modeling. While [Fe/H] is generally not significantly affected, abundances of other elements (e.g. O, S, and K) can differ strongly. The significance of these effects has led us to look to literature sources for NLTE corrections to ensure that our abundances are as accurate as possible (see Section \ref{NLTE} for details.)

In the following section, we provide a review of the literature discussing the importance of the elements we have chosen to analyze. In Section \ref{Data Collection}, we outline our archival data collection process. In Section \ref{Abunance Calculations}, we describe our analysis process. In Section \ref{Results}, we present our results and a comparison with the literature. Lastly, in Section \ref{Discussion}, we provide a discussion on the implications of our results on the planets around our target stars.

\section{Literature Review on Important Elements}\label{LitReview}

\subsection{Fe: The Foundation}

Metallicity (as measured by [Fe/H]) plays an important role in planet formation via the well-known planet-metallicty correlation \citep[e.g.][]{Fischer2005,Wang2015}. Furthermore, \citet{Thorngren+2016} show a strong correlation between the mass of a giant planet and its heavy element enrichment compared to its host star. Specifically, the mass fraction of heavy elements ($Z$) is described by the equation $Z_{planet}/Z_{star} = (9.7 \pm 1.28)M^{(-0.45 \pm 0.09)}$, 
which is approximately
\begin{equation*}
    \frac{Z_{planet}}{Z_{star}} \approx \frac{10}{\sqrt{M_{planet}}}
\end{equation*}

where the mass of the planet is in units of the mass of Jupiter. This relation theoretically allows for an estimate of the metallicity of a giant planet merely from stellar radial velocity (RV) measurements, with no planetary spectroscopy required. This follows from the fact that stellar RVs can be used to derive the mass of the planet, which can then be inserted into the above formula to give an approximate metallicity relationship between the planet and its star.

The iron content of a star also has an influence on the types of planetary systems it can support. \citet{Brewer+2018} show that among planet-hosting stars, those with metallicities below [Fe/H] = -0.3 show an increasing likelihood of hosting compact multi-planet systems (defined as a system with $\geq 3$  planets orbiting $\leq 1$ AU from the host star) relative to the average star of higher metallicity. This trend peaks at [Fe/H] = -0.5, where planet-hosting stars of this metallicity form compact multi-planet systems with triple the frequency of solar-metallicity planet hosts. These compact systems should be easily detectable via the transit method, given the increased frequency of transits of close-in planets.

In terrestrial planets, iron is important for estimations of the core mass fraction (CMF), the ratio of the mass of a planet's core to its total mass. On earth, iron is distributed largely in the core, where it makes up 82.8\% $\pm$ 2.9 of the composition by mass, while it makes up just 6.32\% $\pm$ 0.06 by mass of the mantle \citep{Wang+2018}.

This compositional distinction allows for a simple 2-layer, first-order approximation of a rocky planet, where the core is pure iron, and the mantle is iron-free, being composed of purely magnesium and silicon oxides. This allows for a simplified calculation such that $\textrm{CMF} = M_{Fe}/M_{planet}$, and $M_{planet} = M_{Fe} + M_{SiO_2} + M_{MgO}$. 

\citet{Schulze+2021} use this approximation to explore whether the measured CMF of a planet from its density is consistent with the expected CMF, which is based on the Fe abundance, in conjunction with Mg and Si abundances, of its host star. The results of this paper found that $>$90\% of planets studied show consistent composition, and thus expected CMF, with their host star (see Section \ref{MgSi} for more details). Thus, measurements of stellar Fe abundances, in tandem with Mg and Si as outlined in the following section, can give important clues into the overall composition of terrestrial planets due to the relationship between stellar and planetary compositions.

\subsection{Mg and Si: Rocky Planet Essentials, Atomic Absorbers in Ultra-Hot Jupiters}\label{MgSi}
Together with oxygen, which bonds these elements together in silicate compounds, magnesium and silicon combine to make up 88\% of the earth's mantle \citep{Wang+2018}. This large fractional composition means that getting Mg and Si abundances is of utmost importance when looking to characterize terrestrial planets. 

\citet{Schulze+2021} define three distinct classifications of terrestrial planets based on the ratio $\textrm{CMF}_\rho / \textrm{CMF}_{star}$, where $\textrm{CMF}_\rho$ is a planet's CMF derived from density measurements and $\textrm{CMF}_{star}$ is the same but derived from stellar abundances of Fe, Mg, and Si (for reference, values of this ratio included in the paper for Mercury, Earth, and Mars and are $\sim$2, 1.03, and $\sim$0.6, respectively.) The measurement of $\textrm{CMF}_\rho / \textrm{CMF}_{star}$ requires not just planetary mass and radius constraints, but also constraints on the Mg and Si abundances of a host star relative to its iron abundance.

Planets between $0.5 < \textrm{CMF}_\rho / \textrm{CMF}_{star} < 1.4$ are classified as indistinguishable from their host star in terms of composition, given current uncertainties on mass and radius measurements. Planets with $\textrm{CMF}_\rho / \textrm{CMF}_{star} > 1.4$ are classified as high-density, iron-rich Super-Mercuries, which are assumed to have much larger cores than expected. 

Lastly, planets with $\textrm{CMF}_\rho / \textrm{CMF}_{star} < 0.5$ are classified as "Low-Density Small Planets" (LDSPs). These are distinct from the so-called "super puffs," which are planets with sub-Neptune masses but with transit radii characteristic of gas giants \citep{Wang+2019}. LDSPs are sufficiently dense to indicate a rocky composition, but still have a CMF far lower than expectations. A detailed discussion of LDSPs, with possible explanations for their density deficiencies is presented in Sections 6 and 6.2 of \citet{Schulze+2021}.

% There are multiple compositional variations to explain a core mass deficiency consistent with the definition of an LDSP. For example, one possibility is that a significant amount of oxidized iron is present in the mantle \citep[e.g.][]{Rogers+Seager2010}. Another scenario is that the planet formed from lighter calcium- and aluminum-rich solids which condense early in the life of a protoplanetary disk \citep[e.g.][]{Dorn+2019}. The possibility of a terrestrial planet with a thick atmosphere has also been considered as an explanation for a low bulk density \citep[e.g.][]{Ehrenreich+2012}. Finally, it has been shown that a significant molten mantle, or in the extreme case, magma ocean, would serve to inflate the radius of the planet as well \citep[e.g.][]{Bower+2019}.

Turning now to giant planets, \citet{Lothringer+2021} demonstrate the potential of ultra-hot Jupiters (UHJs) to be used for direct measurements of planetary Mg, Si, and other rock-forming elements' abundances via their emission spectra. These extremely hot (T $>$ 2000K) planets have sufficiently high temperature that these elements do not condense into solids, but rather remain as gases which are detectable through planetary spectroscopy. Such a measurement can be combined with abundance measurements of C, O, and other volatile elements, along with adopted compositions of refractory (i.e. "rocky") and volatile (i.e. "icy") planetesimals, to relate these abundances to the ratio of rocky to icy solids accreted by the planet.

This rock to ice ratio can be used to determine where the planet accreted most of its mass. If the ice fraction is higher, then the UHJ likely accreted much of its solids farther from its star, beyond the snow line \citep{Lothringer+2021}, implying some level of migration. The application of this method of formation tracing is limited to giant planets only. Therefore, measuring stellar abundances for Mg and Si offers a unique opportunity to compare to planetary Mg and Si abundance in order to trace the formation history of short-period gas giant planets.

\subsection{C, N, O, and S: Formation Tracers of Gas Giants, Compositional Indicators of Rocky Planets}\label{CNOS}
Because, in the case of Hot Jupiters, Mg and Si are useful for tracing the formation of only the hottest of such planets, gaseous elements remain extremely important for tracing the formation of cooler giants. It has been shown that a planet's C/O ratio can be related to its formation location in the planetary disk. 

Gas giants that accrete most of their atmospheres in the form of gases beyond the water ice line will have super-stellar C/O. On the other hand, giant planets enriched by accreting significant amounts of planetesimals have C/O similar to that of their host star \citep{Oberg+2011}. This follows from the fact that, beyond the water ice line, most oxygen is trapped in solid water ice particles, meaning a larger proportion of carbon is present in the gas in this region.

In a more recent study, \citet{Espinoza+2017} show that a sub-stellar C/O ratio correlates with metal enrichment from planetesimals, which is consistent with the findings of \citet{Oberg+2011} and \citet{Lothringer+2021}, both of which show that the C/O ratio of giant planets is inversely correlated with their heavy-element enrichment.

In a caveat to this correlation, \citet{Booth+2017} show that it is possible for giant planets with super-stellar C/O ratios to be metal-rich. This unique combination of parameters implies that the planet's metals were not accreted from planetesimals, but instead were accreted largely from metal-enriched gas.

This causes a degeneracy in the case where a planet is observed to have a high C/O ratio. Since a high C/O ratio merely correlates with gas accretion, it offers no information as to the composition of the gas itself (metal-poor or metal-rich). In contrast, observation of a low C/O ratio implies significant accretion of solids, which are necessarily metal-rich. Thus, the above-stated correlation between C/O and metal-enrichment is only valid at low C/O.

However, more detailed constraints can be placed on planet formation by considering other elements as well. Namely, nitrogen and sulfur can be used to differentiate between multiple formation scenarios that produce similar C/O ratios. This is done in \citet{Turrini+2021} by using the planetary abundance ratios N/O, C/N, and S/N, and comparing them to the values of these ratios in the planet's host star. The paper defines the metric

\begin{equation*}
\textrm{X}/\textrm{Y}^* = (\textrm{X}_{planet}/\textrm{Y}_{planet})/(\textrm{X}_{star}/\textrm{Y}_{star})
\end{equation*}

where X and Y are numerical abundances of a given species (e.g. $\textrm{X} = 10^{log(\epsilon_\textrm{x})}$.)\footnote{$log(\epsilon_\textrm{x})$ is the logarithmic abundance of element X, such that $log(\epsilon_\textrm{H})$ is normalized to 12}\footnote{For example, if, for a given star-planet system, X/Y* = 0.5, then the planetary X/Y ratio is equal to half that of the star.} Measuring this quantity in simulated planetary systems allows for constraints on the initial giant planet formation location and the extent of migration more stringent than past studies which have used $\textrm{C}/\textrm{O}$ alone \citep[e.g.][]{Oberg+2011,Mordasini+2016,Madhusudhan+2017}.

\citet{Turrini+2021} show that gas-dominated giant planets have a characteristic abundance pattern of $\textrm{N}/\textrm{O}^* > \textrm{C}/\textrm{O}^* > \textrm{C}/\textrm{N}^*$, whereas solid-enriched giants will instead be characterized by the reverse pattern: $\textrm{C}/\textrm{N}^* > \textrm{C}/\textrm{O}^* > \textrm{N}/\textrm{O}^*$. The size of the spread between these values also correlates with migration of the planet from its initial point of formation, with larger spreads being associated with larger levels of migration.

This is due to the relative positions of the ice lines of compounds of C, N, and O. The ice line of $\textrm{N}_2$, the main source of nitrogen in a planetary disk, is much further from the star than the ice lines of C- and O-carrying compounds (e.g. $\textrm{H}_2\textrm{O}$ and $\textrm{CO}_2$.) This means that gas-dominated giants formed further from the star will contain higher quantities of gaseous nitrogen relative to the amount of gaseous carbon and oxygen, the bulk of which will have condensed into solids at this distance. This would serve to decrease the planetary $\textrm{C}/\textrm{N}^*$ and raise $\textrm{N}/\textrm{O}^*$. The opposite is true for solid-enriched giants, where the relatively carbon- and oxygen-rich solids will boost the planetary $\textrm{C}/\textrm{N}^*$ and lower $\textrm{N}/\textrm{O}^*$.

Furthermore, the $\textrm{S}/\textrm{N}^*$ ratio can be used to constrain the source of the accreted heavy elements in a giant planet. Since sulfur begins to condense into solids closer to the star than does carbon, if the heavy elements are sourced largely from planetesimals, then $\textrm{S}/\textrm{N}^* > \textrm{C}/\textrm{N}^*$. For planets which accreted their heavy elements mainly from enriched nebular gases, $\textrm{C}/\textrm{N}^* > \textrm{S}/\textrm{N}^*$, with greater difference between the values correlating with lesser fractions of solid enrichment.

Clues as to the extent of a giant planet's migration allow for insight into the total mass of planetesimals it has accreted throughout its formation. Simulations by \citet{Shibata+2020} show that a Jupiter-mass planet is capable of accreting roughly 30\% of the planetesimals in its area of influence. For it to accrete further heavy-element mass, migration must be introduced into the model.

This migration allows the planet to leave its current (relatively planetesimal-depleted) orbit and pass through untapped ranges of heavy element material as it makes its way toward its final orbit. In total, such a planet could collect as much as 40-50 earth masses worth of heavy elements by the time it reaches its final orbit. A natural theory to follow would be that Hot Jupiters which have migrated significant distances should deplete the planetesimal resource reservoir for terrestrial planets over a large swath of the proto-planetary disk, making rocky planets unlikely \citep[e.g.][and references therein]{Spalding2017}.

However, \citet{Fogg2007} find that while traveling along their migration path, giant planets leave behind $>$60\% of planetesimals by scattering them to orbits either internal or external to the planet. This percentage of remaining planetesimals is consistent with the $\sim$30\% accretion percentage found by \citet{Shibata+2020}, and still allows for the formation of terrestrial planets. 

This is shown in \citet{Fogg2007} by extending simulations of a planetary disk past the conclusion of the giant planet's migration. Their simulations resulted in the formation of a super-Earth-sized planet outside the final orbit of the giant planet. This rocky planet is rich in volatiles as well, which were shepherded inward by the gas giant, meaning there is potential that such a planet contains significant quantities of water.

These terrestrial planets, outwards of the orbit of their gas-giant counterparts, will tend to enter an orbital resonance which causes their orbits to be tilted relative to the that of the giant planet, making them impossible to detect via transits. Furthermore, other orbital interactions may lead to the outer planet being ejected from the system entirely \citep{Spalding2017}.

The C/O ratio of the host star also has indications for the carbon-richness of terrestrial planets \citep[e.g.][]{Moriarty+2014}. Their simulations show that carbon-rich planets can form in the inner regions of planetary disks ($a \lesssim 0.5~\textrm{AU}$) around stars with C/O as low as 0.65, and can occur throughout the disk around stars with C/O $>$ 0.8.

\subsection{Na and K: Alkali Metals in Giant Planet Atmospheres}
The sodium doublet at 5889\AA~ and 5895\AA, as well as the potassium line at 7698\AA~ should be the most easily detectable features in the spectrum of a Hot Jupiter, largely due to the strength of these lines based on model planetary atmospheres \citep{SeagerSasselov2000}. Indeed, studies show clear detection of sodium \citep[e.g.][and references therein]{Chen+2020}, and use it to get pressure-temperature profiles from the shape of the line \citep[e.g.][HD 209458b]{VidalMadjar+2011}, and sodium abundance from its overall strength \citep[e.g.][WASP-96b]{Nikolov+2018}. While the current precision of planetary sodium abundance measurements is low ($\sigma \simeq 0.5~\textrm{dex}$), future observations with higher precision instruments \citep[e.g. JWST/NIRSpec][]{JWST} will be able to further constrain these values. The sodium abundance can then be compared with that of the planet's host star as a way to directly measure the metal enrichment of the planet compared to its star.

Similar techniques can also be applied to the potassium absorption line \citep[e.g.][]{Chen+2020}, though ground-based observations are more difficult than for sodium due to significant telluric O\textsubscript{2} absorption masking much of the blue half of the line feature \citep{Sedaghati+2016}.

\section{Data Collection}\label{Data Collection}
Our targets were chosen from JWST Cycle 1 GTO and GO programs as outlined in Table \ref{table:JWSTinfo}.

\begin{deluxetable*}{ccc}
\tablecaption{JWST Observation Info\label{table:JWSTinfo}}
\tabletypesize{\footnotesize}
\tablenum{1}
\tablehead{
\hline
\colhead{Star Name} &
\colhead{PI \& Prop. ID} &
\colhead{Science Goal}
}

\startdata
18 Eridani & Beichmann (GTO 1193) & Planet Search/Debris disk characterization \\
18 Indi & Pierre-Olivier Lagage (GTO 1278) & Brown dwarf imaging and spectroscopy \\
55 Cancri & Renyu Hu (GO 1952) & Super-Earth emission spectroscopy \\
HAT-P-1 & David Lafreniere (GTO 1201) & Hot Jupiter transmission and emission spectroscopy\\
HAT-P-26 & Nikole Lewis (GTO 1312) & Hot Jupiter transmission and emission spectroscopy \\
HD 80606 & Tiffany Kataria (GO 2008) & Super-Jupiter phase curve observation \\     
HD 149026 & Johnathon Lunine (GTO 1274) & Hot Jupiter emission spectroscopy \\
HD 189733 & Johnathon Lunine (GTO 1274) & Hot Jupiter emission spectroscopy \\
HD 209458 & Johnathon Lunine (GTO 1274) & Hot Jupiter emission spectroscopy \\
Kepler-51 & Peter Gao (GO 2454) & Super-Puff transmission spectroscopy \\
TOI 193 & David Lafreniere (GTO 1201) & Hot Neptune phase curve observation\\
TOI 421 & Eliza Kempton (GO 1935) & Sub-Neptune transmission spectroscopy \\
WASP-17 & Nikole Lewis (GTO 1353) & Hot Jupiter transmission and emission spectroscopy \\
WASP-52 & David Lafreniere (GTO 1201) & Hot Jupiter transmission spectroscopy\\
WASP-63  & Nestor Espinoza (GO 2113) & Hot Jupiter transmission spectroscopy \\
WASP-77A  & Johnathon Lunine (GTO 1274) & Hot Jupiter emission spectroscopy \\
WASP-127  & David Lafreniere (GTO 1201) & Hot Jupiter transmission spectroscopy \\
\enddata
\end{deluxetable*}

We used archival data from high-resolution optical spectrographs with red wavelength coverage out to at least 7800\AA, and in many cases $>$9200\AA, so that sufficient amount of nitrogen and sulfur features are included, as well as the O I triplet at 7770\AA~and the potassium line at 7698\AA. In order of preference (based on highest sensitivity at red wavelengths), data were sourced from CFHT/ESPaDOnS (3700\AA-10500\AA, R = 68,000), CAHA/CARMENES (5200\AA-9600\AA, R = 94,600), ESO/FEROS (3200\AA-9200\AA, R = 48,000), Keck/HIRES (3360\AA-8100\AA, R = 67,000), and ESO/ESPRESSO (3800\AA-7800\AA, R = 140,000). 

In total, 17 FGK stars had at least one usable spectrum from one of the above instruments. Where possible, multiple exposures were stacked to increase the SNR (which we define as the mean flux divided by the standard flux deviation in a line-feature-less range near the O I triplet) of the final spectrum to at least 100. The number stacked depended on a number of factors, including quantity and quality of spectra available, and thus, this SNR=100 target is not always reached. In the cases of Kepler-51 and WASP-52, we took all the available spectra to produce the highest SNR possible.
Information about the spectra sourced, including SNR, source instrument, and original PI, can be found in Table \ref{table:SpectrumInfo}.

\begin{deluxetable*}{|ccccc|}
\tablecaption{Sources of Stellar Spectra\label{table:SpectrumInfo}}
\tabletypesize{\footnotesize}
\tablenum{2}
\tablehead{
\hline
\colhead{Star Name} &
\colhead{Instrument} &
\colhead{PI and Obs. Date} &
\colhead{\# Stacked} &
\colhead{SNR}
}

\startdata
18 Eridani & ESPaDOnS & Claire Moutou 2014-02-15 \iffalse 05:31:59.684 \fi & 3 & 184 \\
18 Indi & FEROS & Eric Nielsen 2004-09-23 \iffalse 02:08:09.479 \fi & 8 & 117 \\
55 Cancri & ESPaDOnS & Claire Moutou 2018-01-01 \iffalse 9:04:47.671 \fi & 3 & 100 \\
HAT-P-1 & CARMENES & Guijarro 2018-10-18 \iffalse 00:05:34 \fi & 10 & 143 \\
HAT-P-26 & FEROS & Sergio Sousa 2013-01-31 \iffalse 08:49:57.843 \fi & 7 & 130 \\
HD 80606 & HIRES & Stassun 2011-03-15 \iffalse 08:33:54.84 \fi & 2 & 111 \\     
HD 149026 & ESPaDOnS & Gaidos 2012-03-30 \iffalse 12:11:05 \fi & 6 & 305 \\
HD 189733 & ESPaDOnS & Claire Moutou 2013-11-23 \iffalse 4:45:03.980 \fi & 5 & 265 \\
HD 209458 & ESPaDOnS & Claire Moutou 2015-11-29 \iffalse 4:59:28.024 \fi & 4 & 316 \\
Kepler-51 & HIRES & Bedell 2017-08-18 \iffalse 07:57:55.810 \fi & 2 & 53 \\
TOI 193 & ESPRESSO & J.S. Jenkins 2019-11-03 \iffalse 01:55:43.323 \fi & 1 & 102 \\
TOI 421 & HIRES & Howard 2019-09-17 \iffalse 14:37:45.98 \fi & 2 & 114 \\
WASP-17 & FEROS & Francesca Faedi 2012-09-21 \iffalse 00:41:13.096 \fi & 4 & 139 \\
WASP-52 & FEROS & Paula Sarkis 2017-06-09 \iffalse 08:20:04.900 \fi & 3 & 65 \\
WASP-63 & FEROS & Luigi Mancini 2015-01-11 \iffalse 07:10:41.183 \fi & 4 & 157 \\
WASP-77A & FEROS & Luigi Mancini 2014-12-06 \iffalse 01:54:42.283 \fi & 3 & 116 \\
WASP-127 & CARMENES & Fernandez 2019-03-14 \iffalse 00:06:57 \fi & 10 & 114 \\
\enddata
\tablecomments{Spectra for each star were chosen from a single observing run. Thus, for compactness, only a single date of observation is given for each star.}
\end{deluxetable*}

\section{Calculations of Elemental Abundances}\label{Abunance Calculations}

\subsection{Line List}
The iron line list is identical to that used by \citet{Kolecki+2021}. For other elements, we took line data (wavelength, excitation potential, log(gf)) from the NIST database \citep{NIST_ASD}. We made use of this database because of the large quantity of lines of interest for this work, which have been conveniently compiled into a single source. Given that uncertainties in this database vary greatly, we chose to limit our choice of lines to those which had a transition strength accuracy value of 'C' or better. This limits uncertainty in transition strength to a maximum of 25\%. Table \ref{table:LineInfo} lists all lines used in analysis of at least one star in our sample.

\subsection{Removing Telluric-Contaminated Lines}
\begin{figure}[ht]\centering
 \includegraphics[width=\columnwidth]{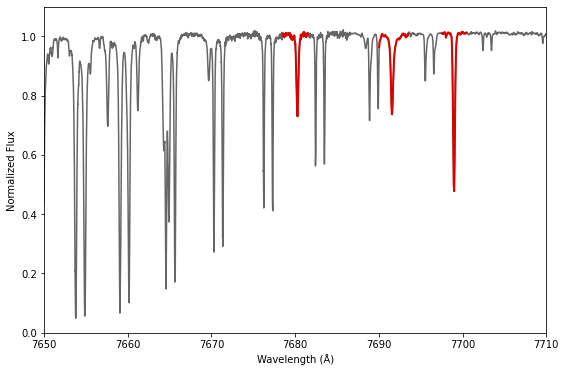}
 \caption{An excerpt from the spectrum of HD 209458 in the range of the red edge of the telluric O\textsubscript{2} A band (seen here as repeating doublets across the spectrum decreasing in depth from left to right). Highlighted in red are three lines of Si, Mg, and K (from left to right), which were used in our analysis. While the left end of the Mg line (center) overlaps with one of the telluric features, this blending is easily ignored by measuring the equivalent width of a model line profile.}\label{fig:tellurics}
\end{figure}

The reduced archival data were not corrected for telluric line contamination. To circumvent this, our line list was chosen to avoid wavelength ranges affected by densely packed telluric molecular bands, most notably the O\textsubscript{2} A and B bands. In wavelength ranges where telluric features are prominent but scattered, we manually ensured that none of the lines chosen for analysis were severely blended with these features. Lines which were only blended at the very edge of the feature were kept, as the blending could easily be neglected by fitting a model line profile. Every line analyzed for each star was visually inspected for blending effects and was discarded if the blending feature could not be removed via modelling. See Figure \ref{fig:tellurics} for a representation of cleanly detected lines within telluric-affected regions.

\subsection{LTE Analysis}
Our abundance analysis follows the methods of \citet{Kolecki+2021}, using equivalent widths with \texttt{abfind} in PyMOOGi \citep{pyMOOGi} to derive elemental abundances. Some changes were made, as outlined in this section.

\subsubsection{Stellar Parameters}

The \citet{Kolecki+2021} parameter fitting is changed, with the addition of UBVRI photometry compiled by SIMBAD to the previously used Gaia, 2MASS, and WISE data \citep[][respectively]{Gaia2018,2MASS,WISE}. Additionally, we took a different approach to sampling stellar parameters, calculating a grid of reduced $\chi^2$ statistics containing each point on an isochrone, from which we derive a probability distribution to directly sample \teff~ and log(g) from. This method produces similar results at that used previously (with differences of roughly 1-sigma or less), but allows for a more robust calculation of the uncertainties.

Also, in the case of non-convergence of microturbulence, rather than setting to 1.5 km/s, we chose to set the value to the point on the microturbulence grid that minimized the magnitude of the slope of the correlation with equivalent width.

\subsubsection{Uncertainty Analysis}\label{Uncertainty}
The uncertainty of abundance measurements is calculated as a quadrature sum of the following sources of error: line-to-line scatter of best fit abundance, and the change of abundance from perturbation of the stellar parameters (\teff, log(g), $\xi$, [Fe/H]) each by 1-$\sigma$. We iterated this calculation to account for the coupled nature of the uncertainties of abundances and stellar parameters. This coupling comes from the fact that the stellar parameters are sampled from isochrones, which vary with metallicity. Thus, a higher uncertainty on [Fe/H] creates a wider distribution of available \teff~and log(g) values to sample from.

Our updated method differs from the previous method \citep{Kolecki+2021,Epstein+2010}. This is because the equilibrium conditions \citep[described in Section 6.1 of][]{Kolecki+2021} are not always met. For example, in this particular work, only 8 stars out of 17 analyzed meet all the conditions to within $2\sigma$. For these particular stars, the two methods produce comparable uncertainties. Therefore, the new uncertainty analysis method builds on our old method and remains valid even in the case of stellar disequilibrium.

Note that in the cases of nitrogen and sodium, additional error terms were introduced into the quadrature sum after NLTE corrections were applied (see Sections \ref{N} and \ref{Na}.)

Finally, uncertainties of numerical abundance ratios were calculated according to the formula for combining multiplied uncertainties

\begin{equation*}
    \sigma_{\textrm{X}/\textrm{Y}} = \frac{\textrm{X}}{\textrm{Y}}*\sqrt{(\frac{\Delta\textrm{X}}{\textrm{X}})^2 + (\frac{\Delta\textrm{Y}}{\textrm{Y}})^2}
\end{equation*}

where $\Delta\textrm{X}$ is the change in numerical abundance by perturbing log($\epsilon_{\textrm{X}}$) by its uncertainty.

\subsection{NLTE corrections}\label{NLTE}
Calculating abundances in LTE sacrifices accuracy for a significant decrease in computational complexity, as many of the simplifying approximations made in the LTE assumption are imperfect representations of the happenings in a stellar interior \citep{Asplund2005}.

These NLTE corrections can be important for interpreting and predicting planet formation pathways. For example, \citet{brewer+2016}'s LTE analysis of 55 Cancri, presented in \citet{BreweCO}, derives a C/O abundance ratio of 0.53. However, \citet{Teske+2013} perform a similar LTE analysis, but with NLTE corrections applied, and derive a C/O ratio of 0.78. \citet{brewer+2016} actually mention the tendency of LTE to overestimate oxygen abundances derived from the O I triplet, which is used in both this work and \citet{brewer+2016}, which would thus underestimate C/O. However, they do not apply any corrections for these effects. The resultant discrepancy is significant beyond the error bars, and also has implications for the composition of 55 Cnc e, as \citet{Moriarty+2014} define the threshold for carbon-rich exoplanet formation at C/O = 0.65.

Several previous papers included calculated abundances with NLTE radiative transfer code \citep[e.g.  MULTI3D,][]{MULTI} for certain sets of stellar parameters. We can thus use these results to improve the accuracy of our analysis. These papers have published values of NLTE corrections, which are the differences in abundances between their NLTE analysis and reference LTE analysis. Adding these differences into our LTE abundances allows us to account for the effects of NLTE without performing the intense calculations usually associated with this. Our NLTE corrections are compiled from a variety of literature sources as listed below. In the event that multiple references for the same corrections were found, we took the most recent available to account for advancements in atomic and atmospheric modelling, as well as in radiative transfer codes.

\subsubsection{Carbon And Oxygen}
For carbon and oxygen, we interpolated the grid of corrections provided by \citet{Amarsi+2019}, which covers the extent of the parameter space of our sample. The magnitude of the carbon corrections was $<$ 0.05 dex for all stars. However, the oxygen corrections were more significant, having an average value of roughly -0.2 dex. In the extreme case of the relatively hot, metal-rich sub-giant HD 149026, the oxygen correction reaches a value of -0.33 dex.

\subsubsection{Nitrogen}\label{N}
For nitrogen, we used the temperature-dependent NLTE corrections of \citet{TakedaHonda2005} by performing a linear regression on their data. This introduced an error of 0.015 dex from the scatter of the data points around the best fit line, which was added in quadrature to the uncertainty from Section \ref{Uncertainty}.

\subsubsection{Sulfur}
For sulfur, we interpolated the grid of \citet{Korotin+2017}. Because the sulfur grid was only calculated for log(g) = 4.0, we attempted to extrapolate values for higher surface gravities. Due to the high excitation potential of the sulfur lines used, we did this by looking at the general trend of the correction vs. log(g) for high-excitation lines.

\begin{figure}[h]\centering
 \includegraphics[width=\columnwidth]{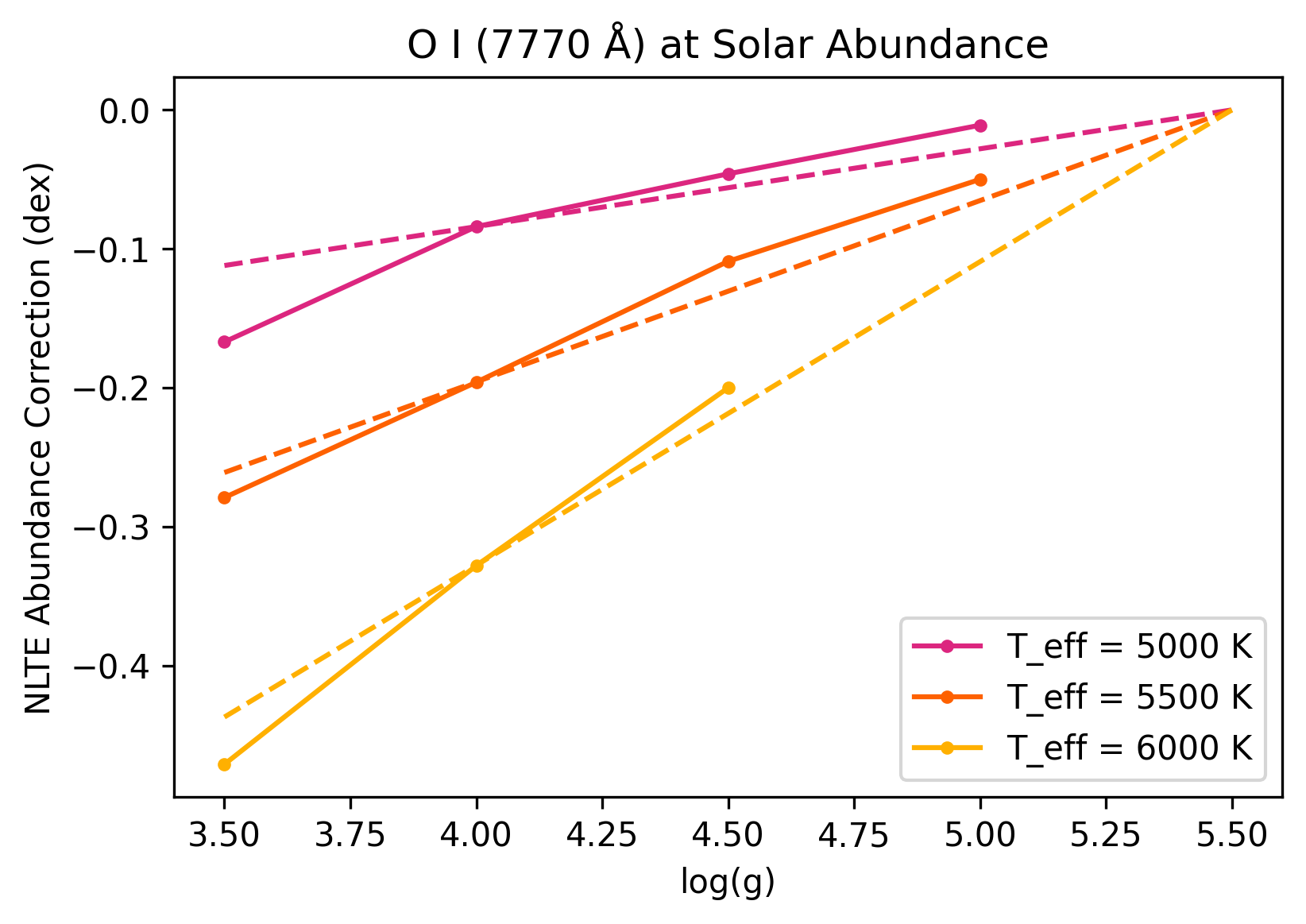}
 \caption{Surface gravity plotted against the magnitude of NLTE corrections for the O I triplet calculated by \citet{Amarsi+2019}. The dashed lines represent an application of our extrapolation method used for sulfur corrections to those of oxygen. The linear extrapolation from log(g) = 4.0 to 5.5 at each temperature is represented by the dashed line, in which the corrections at 5.5 are assumed to be equal to zero. The deviation of the true corrections from our approximation is minimal, $\lesssim$ 0.03 dex in the relevant parameter space.}\label{fig:OINLTE}
\end{figure}

As shown in Figure \ref{fig:OINLTE}, the corrections show a relatively linear trend towards zero as surface gravity increases in the example case of the O I triplet. The physical reasoning for this is as follows. For computational simplicity, the LTE assumption considers all excitations of electrons to be the result of collisions between atoms (local effects). This neglects the contribution of excited electrons caused by incoming radiation from deeper in the star (non-local effects).

This is a largely fair assumption at high pressures (i.e. higher surface gravities), where high pressure implies higher density of particles. This increased density minimizes the mean free path of photons, reducing the impact of non-local radiation. On the other hand, as pressure is lowered and mean free path increases, photon excitation plays a larger role as more and more photons are able to reach the atoms, causing increasing deviations from the LTE assumption at lower surface gravities.

This trend allows for the assumption that at sufficiently high surface gravities, the magnitude of all NLTE corrections approaches zero. We chose to set this critical log(g) value to be 5.5, based on the extrapolations shown in Figure \ref{fig:OINLTE}. To use this point to get sulfur corrections at higher surface gravities and given temperature, we linearly interpolated NLTE corrections along the line which connects the value at log(g) = 4.0 to the point at log(g) = 5.5, where corrections are set to be 0.

These corrections served to lower the sulfur abundance by $<$ 0.1 dex in the case of the 6757\AA~ and 8694\AA~ multiplets. However, the 9212\AA~ multiplet is far more significantly affected. WASP-127 is the star with the most significant sulfur corrections. For this star, the NLTE correction for the 9212\AA~ line is roughly -0.3 dex with respect to LTE.

\subsubsection{Magnesium and Silicon}
For these elements, we used the Spectrum Tools utility \citep[\url{http://nlte.mpia.de},][]{NLTE_MPIA}, which interpolates Mg corrections as calculated for \citet{Bergemann+2017} and Si corrections as calculated for \citet{Bergemann+2013}. These corrections were found to minimally change their respective abundances, lowering them by a maximum of 0.01 dex for our chosen lines.

\subsubsection{Sodium}\label{Na}
For sodium, we adjusted our LTE abundances according to the contour plot presented in Figure 4 of \citet{Lind+2011}. Given our choice of lines, the correction lowers the Na abundance by 0.1 dex for each of our stars. We introduced an uncertainty of $\pm0.025$ dex to account for potential small fluctuations in the correction within the precision of the contour plot. This term was added into the uncertainty from Section \ref{Uncertainty}.

\subsubsection{Potassium}
For potassium, we used an interpolator hosted at \url{http://www2.nao.ac.jp/\textasciitilde takedayi/potassium_nonlte/} \citep{Takeda+2002}. For both WASP-17 and HD 149026, the NLTE correction lowers the LTE potassium abundance by more than 0.5 dex.

\section{Results and Discussion}\label{Results}
Table \ref{table:StellarParams} shows the stellar parameters of our target stars, while Table \ref{table:abs} displays the results of our abundance analysis. Our iron abundances were taken as the average of the abundances derived from Fe I and Fe II lines. To confirm the validity of the comparison to solar abundances, we analyzed a solar spectrum from FEROS and found that our analysis was in agreement with values from \citet{Asplund+2021}. The value of the NLTE corrections applied to the initial LTE abundance, $\textrm{log}(\epsilon_\textrm{x})_{\textrm{LTE}}$, are shown in Table \ref{NLTECorrectionsTable}. In Table \ref{table:Ratios}, we display relevant numerical abundance ratios as mentioned in Sections \ref{MgSi} and \ref{CNOS}.

The empty entries in Table \ref{table:abs} are the result of non-detection of atomic lines of the given species. This is due to the weakness of the lines combined with an SNR which makes them unresolvable from noise.

\begin{deluxetable}{|c c c c c|}[h!]
\tablecaption{Stellar Parameters (units defined as follows: \teff~in K, log(g) in [cm/s\textsuperscript{2}], $\xi$ in km/s, [Fe/H] in dex relative to solar)\label{table:StellarParams}}
%\tabletypesize{\footnotesize}
\tablenum{4}
\tablehead{
\hline
\colhead{Star Name} &
\colhead{\teff} &
\colhead{log(g)} &
\colhead{$\xi$} &
\colhead{[Fe/H]}
}

\startdata
18 Eridani & 5097 $\pm$50 & 4.58 $\pm$0.02 & 0.79 $\pm$0.24 & -0.05 $\pm$0.10 \\
18 Indi & 4682 $\pm$40 & 4.60 $\pm$0.02 & 0.50 $\pm$0.24 & -0.12 $\pm$0.07 \\
55 Cancri & 5308 $\pm$40 & 4.46 $\pm$0.03 & 1.14 $\pm$0.18 & 0.30 $\pm$0.06 \\
HAT-P-1 & 5812 $\pm$190 & 4.26 $\pm$0.09 & 1.16 $\pm$0.07 & 0.01 $\pm$0.11 \\
HAT-P-26 & 5289 $\pm$80 & 4.52 $\pm$0.02 & 0.50 $\pm$0.27 & 0.02 $\pm$0.10 \\
HD 80606 & 5547 $\pm$40 & 4.37 $\pm$0.04 & 1.51 $\pm$0.24 & 0.19 $\pm$0.09 \\    
HD 149026 & 6029 $\pm$20 & 4.20 $\pm$0.02 & 1.04 $\pm$0.11 & 0.31 $\pm$0.09 \\
HD 189733 & 5099 $\pm$30 & 4.56 $\pm$0.02 & 1.52 $\pm$0.18 & -0.12 $\pm$0.08 \\
HD 209458 & 6031 $\pm$20 & 4.31 $\pm$0.02 & 1.12 $\pm$0.13 & -0.01 $\pm$0.06 \\
Kepler-51 & 5577 $\pm$40 & 4.46 $\pm$0.02 & 1.63 $\pm$0.15 & -0.29 $\pm$0.07 \\
TOI 193 & 5410 $\pm$50 & 4.42 $\pm$0.03 & 1.71 $\pm$0.15 & 0.02 $\pm$0.08 \\
TOI 421 & 5324 $\pm$60 & 4.52 $\pm$0.02 & 0.50 $\pm$0.35 & -0.03 $\pm$0.08 \\
WASP-17 & 6157 $\pm$150 & 4.02 $\pm$0.05 & 1.53 $\pm$0.17 & -0.30 $\pm$0.11 \\
WASP-52 & 5121 $\pm$60 & 4.55 $\pm$0.03 & 1.25 $\pm$0.26 & 0.08 $\pm$0.09 \\
WASP-63 & 5512 $\pm$80 & 3.94 $\pm$0.04 & 1.53 $\pm$0.12 & -0.01 $\pm$0.15 \\
WASP-77A & 5660 $\pm$60 & 4.49 $\pm$0.03 & 1.78 $\pm$0.15 & -0.15 $\pm$0.06 \\
WASP-127 & 5949 $\pm$60 & 4.24 $\pm$0.02 & 1.57 $\pm$0.10 & -0.35 $\pm$0.05 \\
\enddata
\tablecomments{$log(\epsilon_{Fe})_\odot = 7.46$ \citep{Asplund+2021}}
\end{deluxetable}

%\begin{longrotatetable}
\begin{deluxetable*}{|cccccccccc|}[h!]
\tabletypesize{\footnotesize}
\tablenum{5}
\tablecaption{Abundance data in log($\epsilon_X$) for target stars\label{table:abs}}
\tablehead{
%\hline
\colhead{Star Name} &
\colhead{Fe} &
\colhead{C} &
\colhead{N} &
\colhead{O} &
\colhead{S} &
\colhead{Mg} &
\colhead{Si} &
\colhead{Na} &
\colhead{K}
}
\startdata
\textbf{Sun}  & \textbf{7.46 $\pm$0.04}  & \textbf{8.46 $\pm$0.04}  &  \textbf{7.83 $\pm$0.07}  &  \textbf{8.69 $\pm$0.04}  &  \textbf{7.12 $\pm$0.03}  &  \textbf{7.55 $\pm$0.03}  &  \textbf{7.51 $\pm$0.03}  &  \textbf{6.22 $\pm$0.03}  &  \textbf{5.07 $\pm$0.03}  \\
18 Eridani & 7.41 $\pm$0.10  & 8.41 $\pm$0.05  &  ... & 8.75 $\pm$0.08 & 6.96 $\pm$0.07 & 7.48 $\pm$0.04  &  7.49 $\pm$0.04 & 6.26 $\pm$0.05 & 5.01 $\pm$0.05 \\
18 Indi & 7.34 $\pm$0.07 & 8.27 $\pm$0.05 & ... & 8.58 $\pm$0.04 & 6.86 $\pm$0.06  &  7.52 $\pm$0.07  &  7.57 $\pm$0.06 & 6.02 $\pm$0.09 & 4.91 $\pm$0.05 \\
55 Cancri & 7.76 $\pm$0.06  & 8.80 $\pm$0.01  & 8.08 $\pm$0.08 &  8.88 $\pm$0.04 & 7.56 $\pm$0.05  &  8.15 $\pm$0.05 & 8.04 $\pm$0.06 & 6.83 $\pm$0.05 & 5.28 $\pm$0.02 \\
HAT-P-1 & 7.47 $\pm$0.11  & 8.49 $\pm$0.09  & 8.05 $\pm$0.09 &  8.78 $\pm$0.07 & 7.08 $\pm$0.10  &  7.67 $\pm$0.05 & 7.70 $\pm$0.04 & 6.33 $\pm$0.08 & 5.07 $\pm$0.08 \\
HAT-P-26 & 7.48 $\pm$0.10  & 8.51 $\pm$0.03  & ... &  8.56 $\pm$0.06 & 7.27 $\pm$0.05  &  7.88 $\pm$0.07 & 7.75 $\pm$0.04 & 6.43 $\pm$0.12 & 5.24 $\pm$0.04 \\
HD 80606 & 7.65 $\pm$0.09  & 8.72 $\pm$0.07  & ... &  8.82 $\pm$0.03 & 7.39 $\pm$0.04  &  8.00 $\pm$0.06 & 7.91 $\pm$0.07 & 6.80 $\pm$0.06 & 5.13 $\pm$0.06 \\
HD 149026 & 7.77 $\pm$0.09  & 8.70 $\pm$0.07  & ... &  8.83 $\pm$0.02 & 7.26 $\pm$0.09  &  7.96 $\pm$0.12 & 7.88 $\pm$0.07 & 6.44 $\pm$0.06 & 5.12 $\pm$0.02 \\
HD 189733 & 7.34 $\pm$0.08  & 8.56 $\pm$0.09  & 7.62 $\pm$0.03 &  8.76 $\pm$0.05 & 6.82 $\pm$0.08  &  7.42 $\pm$0.04 & 7.48 $\pm$0.04 & 6.21 $\pm$0.07 & 5.00 $\pm$0.02 \\
HD 209458 & 7.45 $\pm$0.06  & 8.32 $\pm$0.03  & 7.83 $\pm$0.06 & 8.64 $\pm$0.02 & 7.09 $\pm$0.05  & 7.66 $\pm$0.08 & 7.57 $\pm$0.05 & 6.10 $\pm$0.05 & 4.94 $\pm$0.03 \\
Kepler-51 & 7.17 $\pm$0.07 & 8.23 $\pm$0.23  & ... &  8.68 $\pm$0.04 & ...  &  7.40 $\pm$0.03 & 7.36 $\pm$0.06 & 5.97 $\pm$0.05 & 4.84 $\pm$0.04 \\
TOI 193 & 7.48 $\pm$0.08  & 8.66 $\pm$0.08  & ... &  8.88 $\pm$0.06 & 7.62 $\pm$0.08  &  7.92 $\pm$0.07 & 7.75 $\pm$0.04 & 6.45 $\pm$0.04 & 4.94 $\pm$0.04 \\
TOI 421 & 7.43 $\pm$0.08  & 8.41 $\pm$0.05  & 7.91 $\pm$0.06 &  8.62 $\pm$0.08 & 7.47 $\pm$0.05  &  7.70 $\pm$0.07 & 7.72 $\pm$0.12 & 6.12 $\pm$0.04 & 5.05 $\pm$0.05 \\
WASP-17 & 7.16 $\pm$0.11  & 8.32 $\pm$0.09  & 7.91 $\pm$0.11 &  8.83 $\pm$0.09 & 6.69 $\pm$0.10  &  7.24 $\pm$0.09 & 7.44 $\pm$0.10 & 6.34 $\pm$0.25 & 4.73 $\pm$0.14 \\
WASP-52 & 7.54 $\pm$0.09  & 8.53 $\pm$0.07  & ... &  8.93 $\pm$0.15 & 7.31 $\pm$0.19  &  7.67 $\pm$0.10 & 7.84 $\pm$0.05 & 6.46 $\pm$0.10 & 5.16 $\pm$0.07 \\
WASP-63 & 7.45 $\pm$0.15  & 8.60 $\pm$0.12  & 7.64 $\pm$0.07 &  8.82 $\pm$0.09 & 7.18 $\pm$0.06  &  7.78 $\pm$0.10 & 7.71 $\pm$0.07 & 6.30 $\pm$0.10 & 5.04 $\pm$0.13 \\
WASP-77A & 7.31 $\pm$0.06  & 8.42 $\pm$0.04  & 7.86 $\pm$0.08 &  8.65 $\pm$0.04 & 6.88 $\pm$0.03  &  7.53 $\pm$0.07 & 7.51 $\pm$0.07 & 6.12 $\pm$0.06 & 4.96 $\pm$0.04 \\
WASP-127 & 7.11 $\pm$0.05  & 8.12 $\pm$0.07  & 7.83 $\pm$0.07 &  8.44 $\pm$0.04 & 6.64 $\pm$0.08  &  7.42 $\pm$0.05 & 7.41 $\pm$0.04 & 5.95 $\pm$0.03 & 4.80 $\pm$0.03 \\
\enddata

\tablecomments{Solar abundances are for reference and are sourced from \citet{Asplund+2021}. Null entries in the table are a result of complete non-detection of atomic lines of a given species for a given star.}

\end{deluxetable*}

\begin{deluxetable*}{|cccccccccc|}[h!]
\tabletypesize{\footnotesize}
\tablenum{6}
\tablecaption{NLTE corrections, $\delta$, such that $\textrm{log}(\epsilon_\textrm{x})_{\textrm{LTE}} + \delta = \textrm{log}(\epsilon_\textrm{x})_{\textrm{Final}}$\label{NLTECorrectionsTable}}
\tablehead{
%\hline
\colhead{Star Name} &
\colhead{Fe} &
\colhead{C} &
\colhead{N} &
\colhead{O} &
\colhead{S} &
\colhead{Mg} &
\colhead{Si} &
\colhead{Na} &
\colhead{K}
}
\startdata
18 Eridani & 0.00 & -0.01 & ...   & -0.06 & -0.37 & -0.01 &  0.00 & -0.10 & -0.06\\
18 Indi    & 0.00 & -0.01 & ...   & -0.04 & -0.07 & -0.01 &  0.00 & -0.10 & -0.01\\
55 Cancri  & 0.00 &  0.00 & -0.03 & -0.10 & -0.02 &  0.00 &  0.00 & -0.10 & -0.16\\
HAT-P-1    & 0.00 & -0.03 & -0.07 & -0.24 &  0.00 &  0.00 & -0.01 & -0.10 & -0.33\\
HAT-P-26   & 0.00 &  0.00 & ...   & -0.09 &  0.00 & -0.00 &  0.00 & -0.10 & -0.15\\
HD 80606   & 0.00 & -0.01 & ...   & -0.12 & -0.02 &  0.00 & -0.01 & -0.10 & -0.19\\
HD 149026  & 0.00 &  0.00 & ...   & -0.33 & -0.10 &  0.00 & -0.01 & -0.10 & -0.55\\
HD 189733  & 0.00 & -0.01 & -0.02 & -0.03 & -0.06 & -0.01 & -0.01 & -0.10 & -0.14\\
HD 209458  & 0.00 & -0.02 & -0.09 & -0.27 & -0.09 & -0.01 &  0.00 & -0.10 & -0.50\\
Kepler-51  & 0.00 & -0.02 & ...   & -0.09 & ...   & -0.01 &  0.00 & -0.10 & -0.15\\
TOI 193    & 0.00 &  0.00 & ...   & -0.08 & -0.02 & -0.01 &  0.00 & -0.10 & -0.13\\
TOI 421    & 0.00 & -0.01 & -0.03 & -0.10 & -0.02 & -0.01 &  0.00 & -0.10 & -0.14\\
WASP-17    & 0.00 & -0.03 & -0.10 & -0.34 & -0.06 & -0.01 & -0.01 & -0.10 & -0.62\\
WASP-52    & 0.00 & -0.01 & ...   & -0.05 & -0.07 & -0.01 &  0.00 & -0.10 & -0.11\\
WASP-63    & 0.00 & -0.02 & -0.05 & -0.20 & -0.03 & -0.01 & -0.01 & -0.10 & -0.27\\
WASP-77A   & 0.00 &  0.00 & -0.07 & -0.10 & -0.08 &  0.00 &  0.00 & -0.10 & -0.27\\
WASP-127   & 0.00 & -0.02 & -0.08 & -0.20 & -0.16 & -0.01 & -0.01 & -0.10 & -0.40\\
\enddata
\end{deluxetable*}

%\end{longrotatetable}

\begin{deluxetable*}{|c c c c c c|}[h!]
\tablecaption{Stellar Abundance Ratios\label{table:Ratios}}
%\tabletypesize{\footnotesize}
\tablenum{7}
\tablehead{
\hline
\colhead{Star Name} &
\colhead{C/O} &
\colhead{N/O} &
\colhead{C/N} &
\colhead{S/N} &
\colhead{Mg/Si}
}

\startdata
\textbf{Sun} & \textbf{0.59 $\pm$0.08} & \textbf{0.14 $\pm$0.03} & \textbf{4.27 $\pm$0.85} & \textbf{0.19 $\pm$0.04} & \textbf{1.10 $\pm$0.11} \\
18 Eridani & 0.46 $\pm$0.11 & ... & ... & ... & 0.98 $\pm$0.13\\
18 Indi & 0.49 $\pm$0.08 & ... & ... & ... & 0.89 $\pm$0.20\\
55 Cancri & 0.83 $\pm$0.08 & 0.16 $\pm$0.04 & 5.25 $\pm$1.07 & 0.30 $\pm$0.07 & 1.29 $\pm$0.25\\
HAT-P-1 & 0.51 $\pm$0.17 & 0.19 $\pm$0.06 & 2.75 $\pm$0.85 & 0.11 $\pm$0.03 & 0.93 $\pm$0.13\\
HAT-P-26 & 0.89 $\pm$0.15 & ... & ... & ... & 1.35 $\pm$0.27\\
HD 80606 & 0.79 $\pm$0.15 & ... & ... & ... & 1.23 $\pm$0.28\\    
HD 149026 & 0.74 $\pm$0.13 & ... & ... & ... & 1.20 $\pm$0.44\\
HD 189733 & 0.63 $\pm$0.16 & 0.07 $\pm$0.01 & 8.71 $\pm$2.1 & 0.16 $\pm$0.03 & 0.87 $\pm$0.12\\
HD 209458 & 0.48 $\pm$0.04 & 0.15 $\pm$0.02 & 3.09 $\pm$0.51 & 0.18 $\pm$0.03 & 1.23 $\pm$0.42\\
Kepler-51 & 0.35 $\pm$0.25 & ... & ... & ... & 1.10 $\pm$0.18\\
TOI 193 & 0.60 $\pm$0.15 & ... & ... & ... & 1.48 $\pm$0.30\\
TOI 421 & 0.62 $\pm$0.15 & 0.19 $\pm$0.05 & 3.16 $\pm$0.61 & 0.36 $\pm$0.07 & 0.95 $\pm$0.34\\
WASP-17 & 0.31 $\pm$0.10 & 0.12 $\pm$0.04 & 2.57 $\pm$0.95 & 0.06 $\pm$0.02 & 0.63 $\pm$0.22\\
WASP-52 & 0.40 $\pm$0.18 & ... & ... & ... & 0.68 $\pm$0.19\\
WASP-63 & 0.60 $\pm$0.24 & 0.07 $\pm$0.02 & 9.12 $\pm$3.31 & 0.35 $\pm$0.08 & 1.17 $\pm$0.34\\
WASP-77A & 0.59 $\pm$0.08 & 0.16 $\pm$0.04 & 3.63 $\pm$0.81 & 0.10 $\pm$0.02 & 1.05 $\pm$0.26\\
WASP-127 & 0.48 $\pm$0.10 & 0.25 $\pm$0.05 & 1.95 $\pm$0.48 & 0.06 $\pm$0.02 & 1.02 $\pm$0.16\\
\enddata
\end{deluxetable*}

\subsection{Literature Comparison}\label{litcomp}
We find good agreement among the literature on stellar parameter values (see Table \ref{table:Paramcomp} and Figure \ref{fig:Fe}). Our values are broadly consistent with previously published sources. More discussion can be found in Section \ref{weirdness}. Below, we discuss some targets in more detail.

\subsubsection{HD 189733}

The star that shows the largest discrepancy with Brewer's measurements, HD 189733 ($\feh_{This~work} - \feh_{SPOCS} = -0.18~\textrm{dex}$), has been measured by other works to have a metallicity consistent with our measurement. Past results include \citet[][$\feh = -0.10$]{Montes+2018} and \citet[][$\feh = -0.04$]{Sousa+2018}, both of which fall within $1\sigma$ of the $\feh = -0.12$ derived by this work.

\subsubsection{HAT-P-1}

The other notable abundance discrepancy is that of the [Fe/H] value of HAT-P-1. Our analysis of CARMENES spectra results in [Fe/H] = 0.01, a value much lower than that of \citet{brewer+2016}. We attempted to resolve this discrepancy by analyzing Keck/HIRES data of HAT-P-1 (PI: Asplund 2013-08-16). While this analysis resulted in a higher metallicity than before ([Fe/H] = 0.09), it resulted in an uncharacteristically high sulfur abundance ([S/H] = 0.50). This is in contrast to the abundances of other volatiles in the star, which scaled evenly according to solar metallicity. Furthermore, this higher [Fe/H] was the result of significant ionization disequilibrium, where [Fe I/H] = -0.02 and [Fe II/H] = 0.19. 

It should be noted that the literature distribution of the log(g) value of HAT-P-1 is broad, which may account for some of this variation. The results of this work (log(g) = 4.26) align more closely with the results of \citet{brewer+2016} (log(g) = 4.32) when compared to other literature values \citep[e.g.][log(g) = 4.43]{Liu+2014}.

In the end, we have decided to keep the results of the CARMENES spectrum, which were consistent with reasonable expectations of the distribution of abundances of various elements. Furthermore, the large uncertainties on the stellar parameters of HAT-P-1 in this work (larger than those of other mentioned literature sources) allow for consistency within $2\sigma$ despite otherwise significant deviation.

\subsubsection{WASP-17}

We have also found notable disagreement of \teff~for WASP-17. We derived a \teff~that is significantly lower than that of \citet{Anderson+2010}, who present the discovery of WASP-17~b, by $\sim$400 K. Conversely, our measurement falls within 60 K of that derived by Gaia DR2 \citep{Gaia2018}.

This discrepancy could be due in part to methodology. Gaia effective temperatures are derived from photometry by an algorithm trained on literature catalogs \citep{Andrae+2018}. We also used photometry, though in a different capacity, to derive our stellar parameters. In constrast, \citet{Anderson+2010} use synthetic spectral fitting as in \citet{West+2009}, in which H-alpha and H-beta lines are used to derive effective temperature, and the sodium and magnesium Fraunhofer lines are used to derive surface gravity.

Looking further still, WASP-17 has an extremely broad log(g) distribution, significantly wider than that of HAT-P-1. This suggests that something is off when it comes to getting its stellar parameters, although it is difficult to say for certain what the cause is of this. Spectroscopic surface gravities range from log(g) = 4.14 \citep{Torres+2012} to log(g) = 4.83 \citep{DGM+2015}.

Comparing our results with Gaia again results in fair agreement, as the RV template log(g) = 4.0 for WASP-17. This implies that, albeit on an extremely coarse grid (choosing between logg = 3.5, 4.0, 4.5, 5.0), Gaia’s best spectral fit for the purposes of RV retrieval was a logg of 4.0, which matches up with this work’s value of 4.02.

\subsection{Effects of Analysis Processes on Abundances}\label{weirdness}

\begin{figure*}[htp!]\centering
 \includegraphics[width=\textwidth]{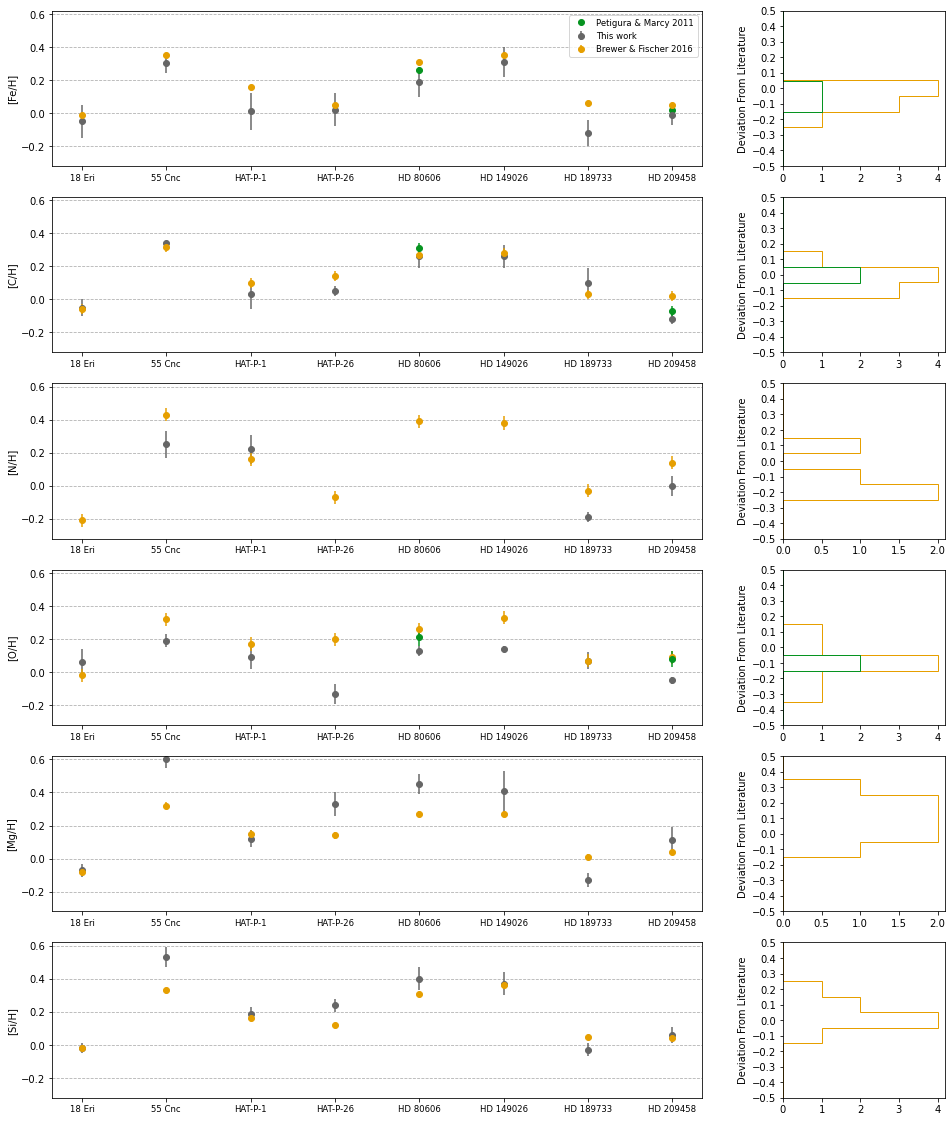}
 \caption{A comparison of abundance values measured by this work (with $1\sigma$ uncertainties) and by \citet{brewer+2016} and \citet{Petigura+Marcy2011}. We found that $> 80\%$ of our measurements fall within 0.1 dex of the literature, which implies good agreement with previously published values, as outlined in Section \ref{weirdness}.}\label{fig:Fe}
\end{figure*}

As is visible in Figure \ref{fig:Fe} and Table \ref{table:Paramcomp}, we found our stellar parameters \teff~and log(g) agree well with those of \citet{brewer+2016}, but our metallicities are systematically lower by 0.08 dex. This could not be explained by differences in the stellar parameters \teff~and log(g), which were minimal by comparison. It also could not be rectified by a simple reanalysis of the sample, which we had hoped would rectify effects of any potential poorly measured line features affecting our results. Thus, we searched the literature for a possible explanation for this discrepancy. 

In general, we found that there can be significant differences in abundance measurements, up to 0.05-0.1 dex \citep{Hinkel+2016}, due to a number of factors. These factors include, but are not limited to: choice of radiative transfer code (e.g. MOOG for this work, SME \citep{SME} for \citet{brewer+2016}), choice of model atmosphere grid (e.g. \citet{ATLAS-APOGEE} for this work, \citet{Castelli2004} for \citet{brewer+2016}), and choice of line list source (e.g. NIST for this work, VALD for \citet{brewer+2016}). See \citet{Hinkel+2016,Jofre+2017,iSpec2019}, and references therein for more detailed discussion of these differences.

Another source of concern was the large scatter of \teff~and logg between this work and the SWEET-Cat catalog \citep{Sousa+2021}. Again looking at Table \ref{table:Paramcomp}, although the mean values of the residuals are approximately 0 (within the error bars), the standard deviation of the residuals of \teff~ is roughly four times the typical error bar on a single \teff~measurement. A similar situation can be seen in the log(g) residuals as well.

Conversely, the residuals of [Fe/H] show a scatter consistent with the measurement error to within 1-2$\sigma$. This implies that our [Fe/H] uncertainties, while larger than others in the literature, are able to account for the systematic errors within the stellar parameters. Therefore, in using the results of this work, it is important to consider the significance of the error bars on our abundance measurements.

While it is not ideal to simply accept these systematic differences, which could significantly affect results, the lack of a single, standardized method for deriving chemical abundances makes it impossible to ensure complete homogeneity of analysis with the literature. Therefore, we have ensured that our abundance analysis process is as rigorous as possible, while acknowledging the potential for minor disagreements, both with and among other rigorously developed pipelines.

With this in mind, we consider our abundances to match up well with the literature, given that more than 80\% of our abunadnces fall within the margin for error discussed in \citet{Hinkel+2016}. The largest deviations are mainly those of nitrogen, an element with extremely difficult-to-measure lines, and of oxygen, which can be explained by our use of NLTE corrections, which lower [O/H] with respect to the values of \citet{brewer+2016}. Other outliers are fairly isolated, and may potentially be resolved by performing multiple analyses of varying methodology to form a distribution of results from which a more robust abundance can be determined.

\subsection{Effects of non-LTE}\label{NLTECO}

\begin{figure}[hb]\centering
 \includegraphics[width=\columnwidth]{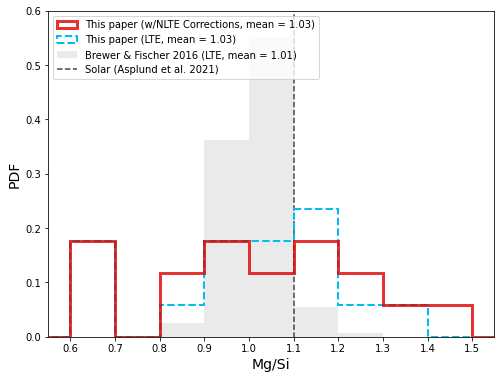}
 \caption{Distribution of Mg/Si ratios of the target stars in LTE and NLTE}\label{fig:MgSi}
\end{figure}

We also compared our resulting distributions of C/O and Mg/Si with those of \citet{BreweCO}. We found good agreement between the average values of our Mg/Si ratios, as shown in Figure \ref{fig:MgSi}, and found that NLTE effects change the Mg/Si derived from an LTE analysis negligibly.

\begin{figure}[hb]\centering
 \includegraphics[width=\columnwidth]{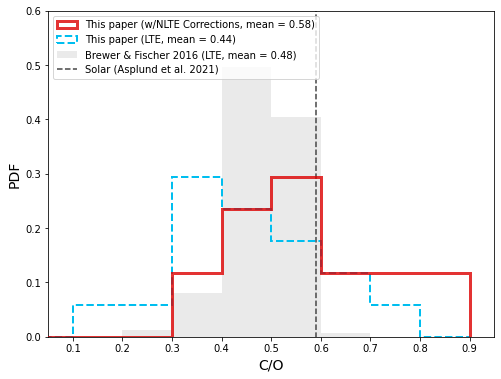}
 \caption{Distribution of C/O ratios of the target stars in LTE and NLTE}\label{fig:CO}
\end{figure}

In contrast, Figure \ref{fig:CO} illustrates the significant NLTE effect on the C/O ratio, as it raises the average of our distribution by 0.14. This is due to the larger decrease in oxygen abundance invoked by NLTE corrections when compared with carbon. \citet{BreweCO} calculate their abundances in LTE without applying NLTE corrections, so we should expect that our mutual LTE C/Os are in agreement, but that our NLTE-corrected C/O ratios are significantly higher. Indeed, we confirm this pattern, also visible in Table \ref{table:COcomp}, which supports the validity of our results.

We discuss that sample selection plays a part in the significant difference in range of values between our two studies in the next section.

\subsection{Discrepancies in Literature C/O Distributions}\label{COdiscrepancies}
In general, we found disagreement among the literature regarding the distribution of C/O in solar-neighborhood planet hosts, largely centered around oxygen abundances. Some studies \citep[e.g.][]{Petigura+Marcy2011} have presented distributions with significantly higher mean values than those of other works \citep[e.g.][]{Nissen2013}, in spite of both papers' efforts to compensate for NLTE effects, which should theoretically increase the accuracy of (and thus decrease discrepancy between) both studies.

\citet{Petigura+Marcy2011} perform spectral synthesis of the forbidden 6300\AA~O I line to derive oxygen abundances, including the blended Ni I feature present in dwarf stars. While this oxygen line is not subject to significant NLTE effects, the blended nickel feature makes an abundance determination via any other method difficult and subject to inaccuracy.

\begin{figure}[ht]\centering
 \includegraphics[width=0.9\columnwidth]{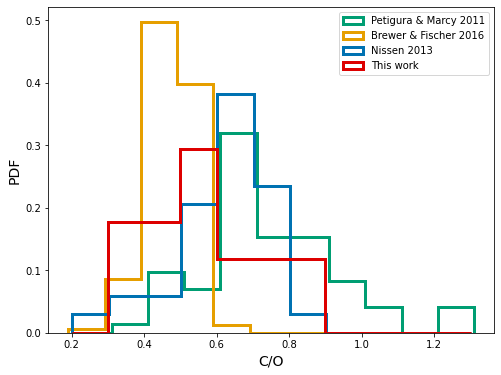}
 \caption{C/O Distribution histograms of this work and various literature sources. The mean values of the distributions are as follows: This work (0.58, LTE w/ NLTE corrections), Brewer \& Fischer (0.48, LTE), Nissen (0.63, LTE w/ NLTE corrections), Petigura \& Marcy (0.76, LTE [O I]).}\label{fig:LitCO}
\end{figure}

\citet{Nissen2013} uses the O I triplet and applies NLTE corrections to an LTE analysis, the same strategy taken by this work. The C/O distribution of \citet{Nissen2013} shows a similar mean and range in Figure \ref{fig:LitCO} as that of this work (Figure \ref{fig:CO}), supporting the validity of our analysis. However, the paper goes on to claim that the detection of significant amounts ($\sim10\%$ of sample stars) of high C/O stars by \citet{Petigura+Marcy2011} is "spurious," attributing this to the difficulty in accurately modeling the 6300\AA~feature with the Ni I blend.

While we do note the significant difference in mean values of the two papers' distribution (0.63 for \citet{Nissen2013} vs 0.76 for \citet{Petigura+Marcy2011}, as shown in Figure \ref{fig:LitCO}), both show a modal peak between 0.6 and 0.7. Further, the spread in values of \citet{BreweCO} is significantly lower than those of the other sources shown in the figure. Thus, a number of other factors beyond methodology could be at play. Firstly, small differences in model atmospheres, NLTE corrections, and spectrum synthesis calculation introduces inherent variability as models differ slightly based on things as seemingly minor as the version of the same radiative transfer code being run \citep[e.g.][and references therein]{Hinkel+2016,Jofre+2017,iSpec2019}. This could serve to introduce results between papers for the same star being analyzed.

Sample size and selection is another factor to consider. We selected 17 planet-hosting targets ($\simeq 0.5\%$ of all such stars discovered at the time of writing\footnote{\url{https://exoplanets.nasa.gov/discovery/exoplanet-catalog/}}), \citet{Nissen2013} selects 33 ($\simeq1\%$), the analysis of \citet{Petigura+Marcy2011} leaves 72 planet hosts with C/O ratios ($\simeq2\%$), and \citet{BreweCO} find the C/O ratios of 163 confirmed hosts ($\simeq4.5\%$). All these studies leave significant gaps in coverage if we are trying to paint a complete picture of all planet-hosting stars. This is especially the case when drawing conclusions about outliers at the extreme high and low ends. 

This low sample size also potentially introduces random selection biases into each individual analysis sample. Perhaps \citet{Petigura+Marcy2011} truly did happen to analyze a subset of the whole with higher-than-usual C/Os. This hypothesis is supported by the two stars in common with our sample in Table \ref{table:COcomp}, which have very similar values from both papers. 

The opposing hypothesis is that the 6300\AA~oxygen feature is a bad indicator of abundance even when compensating for blending effects. The best way to answer this question would be to perform a homogeneous analysis of all the stars in all the samples mentioned, which is beyond the scope of this paper.

In general, however, it seems we should expect a C/O distribution of a large sample of planet hosts, based on either an accurate NLTE analysis of the clean, isolated O I triplet feature, or a joint analysis of the Ni I and [O I] blend, to peak around 0.6 and decrease in either direction in a roughly Gaussian fashion.

\begin{deluxetable}{|c c c c|}[h!]
\tablecaption{Central C/O values derived by this work and by literature sources \citep{Petigura+Marcy2011,BreweCO} The bottom row is the average difference between the reference and this work. \label{table:COcomp}}
%\tabletypesize{\footnotesize}
\tablenum{9}
\tablehead{
\hline
\colhead{Star Name} &
\colhead{This work} &
\colhead{P\&M 2011} &
\colhead{B\&F 2016}
}

\startdata
18 Eridani & 0.46$\pm$0.11 & ... & 0.48 \\
55 Cancri & 0.83$\pm$0.08 &  ... & 0.53 \\
HAT-P-1 & 0.51$\pm$0.17 &  ... & 0.46 \\
HAT-P-26 & 0.89$\pm$0.15 &  ... & 0.46 \\
HD 80606 & 0.79$\pm$0.15 & 0.79$\pm$0.11 & 0.54 \\    
HD 149026 & 0.74$\pm$0.13 & ... & 0.48 \\
HD 189733 & 0.63$\pm$0.16 & ... & 0.49 \\
HD 209458 & 0.48$\pm$0.04 & 0.45$\pm$0.06 & 0.46 \\
\hline
\textbf{Avg. Deviation} &...& \textbf{-0.01} & \textbf{-0.18}
\enddata
\tablecomments{We did not find any common stars between our sample and that of \citet{Nissen2013}}
\end{deluxetable}

\section{Implications}\label{Discussion}

\subsection{Finding Carbon-Rich Worlds}\label{6.1}
Going back to the case of 55 Cancri~e, first mentioned in Section \ref{NLTE}, this is a prime example where NLTE corrections can have a significant impact on abundance results.

With the results of this paper, we can now begin to more precisely estimate the true stellar C/O of 55 Cancri, as we confirm the results of \citet{Teske+2013} (C/O = 0.78), deriving a similarly high C/O ratio in this work of 0.83. Thus, we can see more clearly that 55 Cancri~e is likely a carbon-dominated planet, given that this is well above the \citet{Moriarty+2014} threshold of C/O = 0.65 for short-period carbon-rich planet formation.

This strengthens the case behind the "diamond  planet" model of its interior as discussed in \citet{Madhusudhan+2012}, where it is shown that mass and radius constraints allow for a planetary composition of $> 67\%$ pure carbon, in the form of graphite near the surface, and as diamond as the pressure increases below the surface. This has significant implications as to the potential variations from a purely Earth-like composition of terrestrial exoplanets.

\subsection{Tracking Hot Jupiter Formation}\label{Applying}
A recent paper by \citet{Kawashima+Min2021} provides planetary C/O and N/O for 16 giant planets based on the spectral atmospheric retrieval program ARCiS \citep{Min+2020}. This program uses spectral data to solve for the atmospheric parameters of a planet using disequilibrium chemistry. Five of the planets analyzed by \citet{Kawashima+Min2021} orbit stars analyzed by this work, which allows us to compare our stellar elemental ratios with those computed for the planets.

Using the planetary C/O and N/O ratios, along with C/N (which we calculate for the planets by dividing C/O by N/O), we can use the findings of \citet{Turrini+2021} to inform about the formation pathways for these planets. 

However, we note the large uncertainties associated with the measurement of X/Y*, visible in Table \ref{table:aoipergpoiuharg}. These are largely the result of the propagation of large fractional uncertainties on planetary abundance ratios (often 50\% or higher). We await the science results of JWST for these targets, which should significantly reduce the uncertainties with respect to those presented here. 

Unfortunately this means that, given presently available data and its associated uncertainties, we are unable to draw definitive conclusions on specific cases of planetary formation. Instead, we present possible indications of formation scenarios, which can be verified in the future by higher-precision data.

Given the abundance patterns outlined in Section \ref{CNOS}, we show the possibility that both HD 209458~b and HD 189733~b might have underwent significant migration to get to their present orbits, accreting mostly gas along the way. The higher spread of the X/Y* values of HD 189733~b compared with HD 209458~b may be an indication that HD 189733~b have formed further out from its host star than did HD 209458b. However, we reiterate that the high numerical uncertainties make this far from certain.

HAT-P-1~b, on the other hand, may have formed relatively in situ, close to its star. WASP-17 does not appear to follow either pattern, thus no theories can be posited on its formation based on the data shown here.

\begin{deluxetable}{|c c c c|}[h!]
\tablecaption{Elemental ratios as defined by \citet{Turrini+2021} (see Section \ref{CNOS}) for stars analyzed by this paper with planets analyzed in \citet{Kawashima+Min2021}. In short, values increasing from left to right in the table imply a solid-enriched giant planet, while decreasing values from left to right in the table imply accretion of mostly gaseous matter. Larger spread of these values is associated with higher levels of migration. \label{table:aoipergpoiuharg}}
%\tabletypesize{\footnotesize}
\tablenum{10}
\tablehead{
\hline
\colhead{Planet Name} &
\colhead{N/O*} &
\colhead{C/O*} &
\colhead{C/N*}
}

\startdata
HD 189733 b &  2.14 $\pm$1.32 &  0.75 $\pm$0.41 &  0.36 $\pm$0.29 \\ 
HD 209458 b &  1.27 $\pm$0.62 &  1.04 $\pm$0.53 &  0.85 $\pm$0.52 \\
HAT-P-1 b &  0.79 $\pm$0.52 &  0.80 $\pm$0.69 &  0.99 $\pm$1.04 \\ 
HAT-P-26 b &  ... & 0.28 $\pm$0.14 &  ... \\ 
WASP-17 b &  1.33 $\pm$0.90 &  1.77 $\pm$0.74 &  1.34 $\pm$1.03
\enddata
\end{deluxetable}

\section{Summary}\label{Summary}
In this paper, we have presented a homogeneous abundance analysis of 17 planet-hosting targets. These planets will be observed by JWST during its first observing cycle. Therefore, we are motivated by the prospects of chemical characterization of exoplanets, and how the composition of a planet's host star relates to that of the planet. We present the following as the main conclusions which can be drawn from this work:

\begin{itemize}

    \item Detailed knowledge of a planet's formation requires accurate chemical abundances of its host star. We have detailed the importance of C, N, O, Na, Mg, Si, S, K, and Fe to planet formation in Section \ref{LitReview}.
    
    \item Our abundance analysis produces accurate values for stellar parameters (Table \ref{table:StellarParams}), abundances (Table \ref{table:abs}), and elemental ratios (Table \ref{table:Ratios}) that compare well with the literature where comparisons can be made (see Section \ref{litcomp}). These values can thus be used to inform future studies on planetary composition that take into account host star composition as well. Furthermore, the accuracy of these results supports the use of our updated framework in future studies, which will shed further light onto the composition of planet-hosting stars.
    
    \item NLTE effects on stellar abundance measurements are of strong importance for probing planet formation (see Section \ref{NLTE} and Figure \ref{fig:CO}). Thus, we have compiled sources of corrections for various elements to inform future abundance analyses. We have demonstrated the importance of these corrections with 55 Cancri, a star for which the true C/O ratio has not been fully agreed upon in the literature (see Section \ref{6.1}). NLTE corrections serve to significantly raise the stellar C/O compared to its value calculated in LTE.
    
    \item The precision levels of currently published spectroscopic observations of planets are too low to provide definitive conclusions on the formation and migration of specific giant planets. However, we have shown that possible preliminary conclusions can be drawn (see Section \ref{Applying} and Table \ref{table:aoipergpoiuharg}). Further, as new science results are published from JWST and other upcoming infrared spectroscopic missions, we can expect abundance analyses with significantly higher levels of precision, from which more robust conclusions can be drawn.

\end{itemize}

The targets in this paper are almost entirely gas giant hosts, which is in part due to the SNR requirements for observing planets around FGK stars. Our next paper in this series will focus on M dwarfs, a much larger fraction of which have confirmed planets that are super-Earths or smaller.

\acknowledgments

We would like to thank Chris Sneden for engaging discussion, and his comments and suggestions which have served to improve this work.

This research made use of Astropy,\footnote{\url{http://www.astropy.org}} a community-developed core Python package for Astronomy \citep{astropy:2013, astropy:2018}. This research has made use of the SIMBAD database,
operated at CDS, Strasbourg, France.

Based on observations obtained at the Canada-France-Hawaii 
Telescope (CFHT) which is operated by the National Research Council 
of Canada,the Institut National des Sciences de l'Univers      
of the Centre National de la Recherche Scientique of France,            
and the University of Hawaii. 

This research has made use of the Keck Observatory Archive (KOA), 
which is operated by the W. M. Keck Observatory and the NASA 
Exoplanet Science Institute (NExScI), under contract with the 
National Aeronautics and Space Administration.

Based on data from observations collected at the European Southern Observatory under ESO programs 073.C-0528(A), 090.C-0146(A), 103.2028.001, 089.C-0471(A), 099.A-9010(A), and 094.A-9010(A).

Based on data from the CAHA Archive at CAB (INTA-CSIC).

\software{PyMOOGi \citep{pyMOOGi}, MOOG \citep{Sneden1973}, SciPy \citep{Scipy}, Astropy \citep{AstroPy}}

\bibliography{bib}
\newgeometry{tmargin=0.1cm,bmargin=-1.61in,lmargin = 0.5in, rmargin = 0.5in}
\begin{longrotatetable}
\begin{deluxetable}{|c m{.25in} m{.3in} c | m{.25in} m{.25in} m{.25in} m{.25in} m{.25in} m{.25in} m{.25in} m{.25in} m{.25in} m{.25in} m{.25in} m{.25in} m{.25in} m{.25in} m{.25in} m{.25in} m{.25in}|}
\tablecaption{Line Data\label{table:LineInfo}}
\tabletypesize{\tiny}
\tablenum{3}
\tablehead{
\multicolumn{4}{c}{Line Information} &
\multicolumn{17}{c}{Equivalent Width For Given Star (m\AA)} \cr
\colhead{$\lambda$ (\AA)} &
\colhead{X} &
\colhead{E.P.(eV)} &
\colhead{log(gf)} &
\colhead{18 Eri} &
\colhead{18 Ind} &
\colhead{55 Cnc} &
\colhead{HAT-P-1} &
\colhead{HAT-P-26} &
\colhead{HD 80606} &
\colhead{HD 149026} &
\colhead{HD 189733} &
\colhead{HD 209458} &
\colhead{Kepler-51} &
\colhead{TOI 193} &
\colhead{TOI 421} &
\colhead{WASP-17} &
\colhead{WASP-52} &
\colhead{WASP-63} &
\colhead{WASP-77A} &
\colhead{WASP-127}
}

\startdata
  4932.030 &   C I &  7.68 & -1.66 &      ...  &      ...  &      ...  &      ...  &      ...  &      ...  &    61.75 &      ...  &      ...  &    27.07 &      ...  &      ...  &    39.72 &      ...  &      ...  &      ...  &      ...  \\
  5380.330 &   C I &  7.68 & -1.62 &      ...  &    21.61 &      ...  &    23.07 &    12.79 &    26.69 &    39.49 &    10.75 &    22.03 &      ...  &    18.24 &      ...  &    25.95 &    11.83 &    35.47 &    15.67 &    26.58 \\
  6587.620 &   C I &  8.54 & -1.00 &      ...  &      ...  &      ...  &    20.00  &      ...  &    15.41 &    32.41 &      ...  &    18.05 &      ...  &     9.52 &      ...  &      ...  &      ...  &    16.89 &      ...  &    11.42 \\
  7111.460 &   C I &  8.64 & -1.08 &      ...  &      ...  &      ...  &    11.87 &      ...  &      ...  &    19.76 &      ...  &      ...  &      ...  &    14.40 &      ...  &      ...  &      ...  &    12.47 &    10.73 &     5.58 \\
  7113.170 &   C I &  8.65 & -0.77 &      ...  &      ...  &      ...  &    21.70 &      ...  &      ...  &    59.17 &      ...  &    29.53 &    11.64 &    22.78 &     9.70 &      ...  &     9.34 &    48.98 &    18.10 &    16.66 \\
  7116.980 &   C I &  8.65 & -0.91 &      ...  &      ...  &      ...  &    21.98 &      ...  &      ...  &    37.63 &     9.37 &    21.80 &      ...  &      ...  &      ...  &    25.79 &      ...  &    21.87 &      ...  &    14.81 \\
  8058.620 &   C I &  8.84 & -1.18 &      ...  &      ...  &      ...  &    11.32 &      ...  &      ...  &    14.48 &      ...  &      ...  &      ...  &      ...  &      ...  &      ...  &      ...  &    10.68 &      ...  &      ...  \\
  8335.140 &   C I &  7.68 & -0.42 &      ...  &      ...  &    15.32 &      ...  &      ...  &      ...  &      ...  &      ...  &   102.39 &    37.44 &      ...  &      ...  &   129.20 &    42.06 &    69.73 &      ...  &    69.04 \\
  9061.436 &   C I &  7.48 & -0.34 &      ...  &      ...  &      ...  &      ...  &      ...  &      ...  &      ...  &    53.58 &      ...  &      ...  &      ...  &      ...  &      ...  &      ...  &      ...  &      ...  &      ...  \\
  9094.830 &   C I &  7.48 &  0.14 &      ...  &      ...  &      ...  &      ...  &      ...  &      ...  &      ...  &   126.05 &      ...  &      ...  &      ...  &      ...  &      ...  &      ...  &      ...  &      ...  &      ...  \\
  9111.807 &   C I &  7.48 & -0.34 &    63.13 &      ...  &      ...  &   185.90 &      ...  &      ...  &      ...  &      ...  &      ...  &      ...  &      ...  &      ...  &      ...  &      ...  &      ...  &      ...  &   139.94 \\
  7468.312 &   N I & 10.34 & -0.18 &      ...  &     ... &      ...  &     5.46 &      ...  &     ... &    ... &      ...  &     4.94 &      ...  &     ... &     1.25 &      ...  &     ... &      ...  &     2.48 &             ...  \\
  8184.860 &   N I & 10.33 & -0.30 &      ... &      ...  &     2.30 &      ...  &      ...  &      ...  &      ...  &      ...  &      ...  &      ...  &      ...  &      ...  &     6.12 &      ...  &      ...  &      ...  &      ...  \\
  8188.010 &   N I & 10.33 & -0.30 &      ...  &      ...  &      ...  &      ...  &      ...  &      ...  &    ... &    18.27 &     4.11 &      ...  &      ...  &      ...  &      ...  &      ...  &      ...  &      ...  &     4.20 \\
  8216.340 &   N I & 10.34 &  0.14 &      ...  &      ...  &      ...  &      ...  &     ... &      ...  &      ...  &      ...  &      ...  &      ...  &      ...  &      ...  &    16.87 &     ... &     3.66 &      ...  &      ...  \\
  9392.790 &   N I & 10.69 &  0.30 &      ...  &      ...  &      ...  &      ...  &      ...  &      ...  &      ...  &     8.13 &      ...  &      ...  &      ...  &      ...  &      ...  &      ...  &      ...  &      ...  &     6.29 \\
  7771.940 &   O I &  9.15 &  0.37 &    30.20 &    47.66 &      47.65  &    89.65 &    35.81 &    61.50 &   117.94 &    26.84 &    99.91 &    60.68 &    53.57 &    39.81 &   145.41 &    30.91 &    74.59 &    60.43 &    81.62 \\
  7774.170 &   O I &  9.15 &  0.22 &    25.55 &    43.88 &      43.88  &    80.64 &      ...  &    60.39 &   106.87 &    27.10 &    84.15 &    52.89 &    47.59 &    27.16 &   123.76 &    24.59 &    69.92 &    56.99 &    67.16 \\
  7775.390 &   O I &  9.15 &  0.00 &    14.63 &    32.83 &      32.83  &    64.54 &    18.41 &    41.76 &    85.22 &    17.32 &    68.56 &    35.40 &    43.81 &    25.11 &    98.74 &    36.71 &    56.89 &    41.62 &    48.00 \\
  4668.560 &  Na I &  2.10 & -1.31 &    89.14 &      ...  &    93.30 &      ...  &      ...  &      ...  &      ...  &      ...  &      ...  &      ...  &      ...  &      ...  &      ...  &      ...  &      ...  &      ...  &      ...  \\
  4978.541 &  Na I &  2.10 & -1.22 &      ...  &      ...  &      ...  &      ...  &      ...  &      ...  &      ...  &      ...  &      ...  &      ...  &      ...  &      ...  &    81.89 &      ...  &      ...  &      ...  &      ...  \\
  5682.633 &  Na I &  2.10 & -0.71 &      ...  &      ...  &   238.14 &   134.20 &   238.53 &      ...  &      ...  &      ...  &      ...  &      ...  &      ...  &      ...  &      ...  &      ...  &   128.94 &   113.56 &      ...  \\
  5688.205 &  Na I &  2.10 & -0.45 &      ...  &   298.27 &   242.90 &   142.47 &   187.87 &   198.98 &   144.38 &   200.68 &   116.58 &   128.54 &   223.88 &   179.75 &   114.19 &   251.49 &   135.83 &   139.74 &   101.38 \\
  6154.225 &  Na I &  2.10 & -1.55 &    64.89 &   107.88 &    92.68 &    39.61 &    68.44 &    83.86 &    50.04 &    79.71 &    30.50 &    39.88 &    72.92 &    44.73 &     9.24 &    96.16 &    66.82 &    50.08 &    24.45 \\
  6160.747 &  Na I &  2.10 & -1.25 &    99.62 &      ...  &      ...  &      ...  &      ...  &      ...  &    55.75 &      ...  &    37.99 &    48.85 &      ...  &    68.23 &      ...  &      ...  &    74.71 &    61.20 &    38.69 \\
  8194.790 &  Na I &  2.10 & -0.46 &      ...  &      ...  &      ...  &      ...  &      ...  &      ...  &      ...  &   497.28 &      ...  &      ...  &      ...  &      ...  &   184.86 &      ...  &      ...  &      ...  &      ...  \\
  9961.256 &  Na I &  3.62 & -0.82 &      ...  &    65.20 &      ...  &      ...  &      ...  &      ...  &      ...  &    30.78 &      ...  &      ...  &      ...  &      ...  &      ...  &      ...  &      ...  &      ...  &      ...  \\
  4057.505 &  Mg I &  4.35 & -0.90 &      ...  &      ...  &      ...  &      ...  &      ...  &      ...  &      ...  &      ...  &      ...  &      ...  &      ...  &      ...  &   212.59 &      ...  &      ...  &      ...  &      ...  \\
  5528.405 &  Mg I &  4.35 & -0.50 &      ...  &   576.95 &      ...  &      ...  &   602.71 &   508.13 &   293.59 &      ...  &   302.50 &   436.67 &      ...  &      ...  &   182.00 &   512.54 &   357.16 &   354.95 &   284.47 \\
  7387.689 &  Mg I &  5.75 & -1.00 &    92.86 &   207.01 &    95.97 &    92.51 &   155.92 &   152.73 &   115.16 &    91.75 &      ...  &    83.85 &   140.90 &   110.46 &      ...  &    98.85 &   124.75 &    87.51 &    59.00 \\
  7691.553 &  Mg I &  5.75 & -0.78 &   140.66 &   240.87 &   112.64 &   131.53 &   171.60 &   200.20 &   177.98 &   166.89 &   114.05 &    96.96 &   205.08 &   167.01 &    47.01 &   224.86 &      ...  &   111.51 &    89.19 \\
  8305.596 &  Mg I &  5.93 & -1.32 &    39.72 &      ...  &      ...  &      ...  &      ...  &      ...  &      ...  &      ...  &      ...  &      ...  &      ...  &      ...  &      ...  &      ...  &      ...  &      ...  &      ...  \\
  8736.020 &  Mg I &  5.95 & -0.72 &      ...  &   326.08 &   159.86 &      ...  &   223.15 &   309.13 &   219.73 &   235.08 &      ...  &      ...  &      ...  &      ...  &      ...  &   155.57 &   167.48 &   157.52 &   110.07 \\
  8923.569 &  Mg I &  5.39 & -1.68 &    72.47 &   109.94 &    66.56 &    53.96 &    93.65 &      ...  &    61.03 &    76.38 &    45.61 &      ...  &      ...  &      ...  &    34.07 &    84.80 &    69.69 &    57.39 &    48.39 \\
  9432.764 &  Mg I &  5.93 & -0.92 &    93.54 &      ...  &      ...  &      ...  &      ...  &      ...  &      ...  &      ...  &   117.33 &      ...  &      ...  &      ...  &      ...  &      ...  &      ...  &      ...  &    54.32 \\
  5645.613 &  Si I &  4.93 & -1.63 &      ...  &      ...  &      ...  &      ...  &      ...  &    75.92 &    53.54 &      ...  &    44.04 &    41.50 &      ...  &      ...  &      ...  &      ...  &      ...  &    39.44 &      ...  \\
  5665.555 &  Si I &  4.92 & -2.04 &    39.52 &    76.28 &    26.59 &    44.91 &    53.10 &    62.70 &    56.18 &    43.90 &    36.94 &    37.27 &    65.66 &    63.33 &      ...  &    67.62 &    58.86 &    52.50 &    30.64 \\
  5684.484 &  Si I &  4.95 & -1.42 &    61.71 &      ...  &    46.24 &      ...  &      ...  &      ...  &      ...  &      ...  &      ...  &      ...  &      ...  &      ...  &      ...  &      ...  &      ...  &      ...  &      ...  \\
  5690.425 &  Si I &  4.93 & -1.87 &    42.78 &    72.70 &    31.38 &    56.24 &    47.90 &    65.85 &    62.29 &    42.39 &    44.96 &    61.31 &    64.54 &      ...  &    24.08 &    61.01 &    65.92 &    53.93 &    54.28 \\
  5701.104 &  Si I &  4.93 & -2.05 &    30.73 &    56.88 &      ...  &      ...  &    49.88 &      ...  &      ...  &    33.59 &      ...  &    27.70 &    51.40 &      ...  &      ...  &    48.22 &    55.36 &      ...  &    22.33 \\
  5708.400 &  Si I &  4.95 & -1.47 &      ...  &      ...  &      ...  &      ...  &      ...  &      ...  &      ...  &      ...  &      ...  &      ...  &      ...  &      ...  &    73.69 &      ...  &      ...  &      ...  &      ...  \\
  5772.146 &  Si I &  5.08 & -1.75 &      ...  &      ...  &    37.09 &    52.47 &      ...  &    73.90 &    65.44 &      ...  &    47.29 &    43.76 &    70.45 &    43.72 &    28.66 &    56.42 &    69.88 &    52.99 &    39.08 \\
  5797.856 &  Si I &  4.95 & -2.05 &      ...  &      ...  &      ...  &      ...  &      ...  &      ...  &      ...  &      ...  &      ...  &    26.09 &      ...  &      ...  &    55.27 &      ...  &      ...  &      ...  &      ...  \\
  5948.541 &  Si I &  5.08 & -1.23 &      ...  &   129.14 &    71.89 &    89.97 &   105.81 &   131.18 &   109.47 &    99.89 &    86.65 &    83.71 &   100.51 &    88.95 &    60.20 &   106.32 &   112.52 &    98.65 &    83.24 \\
  7680.266 &  Si I &  5.86 & -0.69 &    66.46 &      ...  &    56.47 &    96.55 &    89.40 &      ...  &   102.13 &    70.29 &    79.93 &      ...  &    92.25 &      ...  &    56.39 &    95.79 &    99.49 &    94.16 &    70.29 \\
  8093.232 &  Si I &  5.86 & -1.35 &      ...  &      ...  &      ...  &      ...  &      ...  &      ...  &      ...  &      ...  &      ...  &      ...  &      ...  &      ...  &      ...  &      ...  &      ...  &      ...  &    28.91 \\
  8648.470 &  Si I &  6.21 &  0.05 &   116.43 &      ...  &   121.18 &   171.65 &   150.97 &   189.23 &   187.29 &   122.47 &   144.69 &      ...  &      ...  &      ...  &   137.25 &      ...  &   121.21 &   135.81 &   111.54 \\
  5706.100 &   S I &  7.87 & -0.93 &      ...  &      ...  &      ...  &      ...  &      ...  &      ...  &      ...  &      ...  &      ...  &    94.32 &      ...  &      ...  &      ...  &      ...  &      ...  &      ...  &      ...  \\
  6757.150 &   S I &  7.87 & -0.35 &      ...  &    15.71 &      ...  &      ...  &      ...  &    22.04 &      ...  &      ...  &    22.13 &      ...  &    19.57 &    12.43 &    33.05 &      ...  &      ...  &     7.93 &    14.34 \\
  6757.180 &   S I &  7.87 & -0.24 &      ...  &      ...  &      ...  &      ...  &      ...  &      ...  &    39.39 &      ...  &      ...  &      ...  &      ...  &      ...  &      ...  &      ...  &      ...  &      ...  &      ...  \\
  8633.120 &   S I &  8.40 & -0.06 &      ...  &      ...  &      ...  &    12.32 &      ...  &      ...  &      ...  &      ...  &      ...  &      ...  &      ...  &      ...  &      ...  &      ...  &      ...  &      ...  &      ...  \\
  8655.170 &   S I &  8.40 & -0.54 &      ...  &      ...  &      ...  &      ...  &     2.18 &      ...  &      ...  &      ...  &      ...  &      ...  &      ...  &      ...  &      ...  &      ...  &      ...  &      ...  &      ...  \\
  8693.980 &   S I &  7.87 & -0.54 &      ...  &      ...  &      ...  &      ...  &      ...  &      ...  &      ...  &     7.95 &    12.69 &      ...  &      ...  &      ...  &      ...  &     8.55 &      ...  &     6.17 &     6.52 \\
  8694.630 &   S I &  7.86 &  0.08 &      ...  &      ...  &      ...  &    32.10 &      ...  &      ...  &    44.18 &      ...  &      ...  &      ...  &      ...  &      ...  &      ...  &      ...  &      ...  &      ...  &      ...  \\
  8694.710 &   S I &  7.87 &  0.05 &     6.49 &    27.44 &      ...  &      ...  &      ...  &      ...  &      ...  &     5.66 &    33.11 &      ...  &      ...  &      ...  &    27.77 &      ...  &    29.51 &      ...  &    19.56 \\
  9212.860 &   S I &  6.52 &  0.42 &      ...  &      ...  &    42.68 &      ...  &      ...  &      ...  &   171.10 &      ...  &      ...  &      ...  &      ...  &      ...  &      ...  &      ...  &      ...  &      ...  &      ...  \\
  9212.863 &   S I &  6.52 &  0.40 &    87.86 &      ...  &      ...  &      ...  &      ...  &      ...  &      ...  &      ...  &   192.76 &      ...  &      ...  &      ...  &      ...  &    89.40 &      ...  &   129.39 &   145.91 \\
  9228.090 &   S I &  6.52 &  0.26 &      ...  &      ...  &      ...  &   107.35 &      ...  &      ...  &      ...  &      ...  &      ...  &      ...  &      ...  &      ...  &      ...  &      ...  &      ...  &      ...  &      ...  \\
  9228.093 &   S I &  6.52 &  0.25 &      ...  &      ...  &      ...  &      ...  &      ...  &      ...  &      ...  &      ...  &      ...  &      ...  &      ...  &      ...  &      ...  &      ...  &      ...  &      ...  &    92.26 \\
  9237.538 &   S I &  6.52 &  0.03 &      ...  &      ...  &      ...  &      ...  &      ...  &      ...  &      ...  &    69.96 &   125.39 &      ...  &      ...  &      ...  &      ...  &      ...  &      ...  &      ...  &    96.54 \\
  7698.965 &   K I &  0.00 & -0.18 &   251.98 &   254.14 &   373.88 &   166.38 &   261.72 &   196.72 &   169.24 &   284.97 &   152.35 &   165.63 &   191.34 &   208.31 &   133.12 &   288.37 &   181.10 &   191.91 &   152.55 \\
  4007.272 &  Fe I &  2.76 & -1.28 &   114.93 &      ...  &      ...  &      ...  &   110.23 &      ...  &      ...  &      ...  &      ...  &      ...  &      ...  &    96.23 &    68.46 &      ...  &      ...  &   105.74 &      ...  \\
  4009.713 &  Fe I &  2.22 & -1.25 &      ...  &   255.35 &      ...  &      ...  &      ...  &      ...  &      ...  &      ...  &      ...  &      ...  &      ...  &      ...  &   105.71 &      ...  &      ...  &      ...  &      ...  \\
  4017.148 &  Fe I &  3.05 & -1.06 &      ...  &      ...  &   149.76 &      ...  &      ...  &      ...  &      ...  &      ...  &      ...  &      ...  &      ...  &      ...  &      ...  &      ...  &      ...  &      ...  &      ...  \\
  4067.271 &  Fe I &  2.56 & -1.42 &   119.24 &      ...  &      ...  &      ...  &   120.32 &      ...  &      ...  &      ...  &      ...  &      ...  &      ...  &      ...  &      ...  &      ...  &      ...  &      ...  &      ...  \\
  4067.978 &  Fe I &  3.21 & -0.47 &      ...  &      ...  &      ...  &      ...  &      ...  &      ...  &      ...  &      ...  &      ...  &      ...  &      ...  &      ...  &   114.53 &      ...  &      ...  &      ...  &      ...  \\
  4072.502 &  Fe I &  3.43 & -1.44 &      ...  &      ...  &      ...  &      ...  &      ...  &      ...  &      ...  &      ...  &      ...  &      ...  &      ...  &      ...  &    51.75 &      ...  &      ...  &      ...  &      ...  \\
  4073.762 &  Fe I &  3.27 & -0.90 &      ...  &   169.69 &      ...  &      ...  &      ...  &      ...  &   107.55 &      ...  &      ...  &      ...  &      ...  &   122.81 &    74.16 &      ...  &      ...  &      ...  &      ...  \\
  4078.353 &  Fe I &  2.61 & -1.47 &      ...  &      ...  &      ...  &      ...  &      ...  &      ...  &      ...  &      ...  &      ...  &      ...  &      ...  &      ...  &    86.11 &      ...  &      ...  &      ...  &      ...  \\
  4079.838 &  Fe I &  2.86 & -1.36 &   116.45 &   166.22 &      ...  &      ...  &      ...  &      ...  &    89.00 &      ...  &    89.18 &   144.79 &      ...  &   130.95 &      ...  &   145.58 &   129.10 &   106.31 &      ...  \\
  4107.488 &  Fe I &  2.83 & -0.88 &      ...  &   187.20 &      ...  &      ...  &      ...  &      ...  &      ...  &      ...  &      ...  &      ...  &      ...  &      ...  &    94.58 &      ...  &   142.00 &   146.44 &      ...  \\
  4120.206 &  Fe I &  2.99 & -1.27 &   135.84 &      ...  &   143.39 &      ...  &   129.47 &      ...  &    95.12 &      ...  &      ...  &      ...  &   133.74 &   121.49 &    67.11 &   130.24 &   108.64 &   105.19 &      ...  \\
  4121.802 &  Fe I &  2.83 & -1.45 &   123.83 &   139.01 &      ...  &      ...  &   106.17 &      ...  &   110.54 &      ...  &    82.50 &      ...  &      ...  &      ...  &    65.43 &      ...  &   107.53 &   100.64 &      ...  \\
  4136.998 &  Fe I &  3.41 & -0.45 &   147.09 &      ...  &      ...  &      ...  &      ...  &      ...  &   125.70 &      ...  &      ...  &   119.03 &      ...  &   136.56 &      ...  &      ...  &   119.11 &      ...  &      ...  \\
  4157.780 &  Fe I &  3.42 & -0.40 &      ...  &      ...  &      ...  &      ...  &      ...  &      ...  &   133.99 &      ...  &      ...  &      ...  &      ...  &      ...  &   100.39 &      ...  &      ...  &      ...  &      ...  \\
  4175.636 &  Fe I &  2.85 & -0.83 &      ...  &      ...  &      ...  &      ...  &      ...  &      ...  &      ...  &      ...  &      ...  &      ...  &      ...  &      ...  &    97.68 &      ...  &      ...  &      ...  &      ...  \\
  4184.891 &  Fe I &  2.83 & -0.87 &      ...  &   188.77 &      ...  &      ...  &   147.74 &      ...  &   119.10 &      ...  &   105.72 &   137.94 &      ...  &   136.41 &      ...  &      ...  &   141.04 &   135.16 &      ...  \\
  4196.209 &  Fe I &  3.40 & -0.70 &      ...  &      ...  &      ...  &      ...  &      ...  &      ...  &    98.04 &      ...  &      ...  &      ...  &      ...  &      ...  &      ...  &      ...  &      ...  &      ...  &      ...  \\
  4219.359 &  Fe I &  3.57 & -0.00 &      ...  &   259.32 &      ...  &      ...  &      ...  &      ...  &      ...  &      ...  &      ...  &      ...  &      ...  &      ...  &      ...  &      ...  &      ...  &      ...  &      ...  \\
  4222.213 &  Fe I &  2.45 & -0.97 &      ...  &      ...  &      ...  &      ...  &      ...  &      ...  &   146.12 &      ...  &   145.81 &      ...  &      ...  &      ...  &   103.62 &      ...  &      ...  &      ...  &      ...  \\
  4224.171 &  Fe I &  3.37 & -0.51 &      ...  &   236.43 &      ...  &      ...  &      ...  &      ...  &      ...  &      ...  &      ...  &      ...  &      ...  &      ...  &      ...  &      ...  &      ...  &      ...  &      ...  \\
  4266.964 &  Fe I &  2.73 & -1.81 &      ...  &      ...  &      ...  &      ...  &      ...  &      ...  &      ...  &      ...  &      ...  &      ...  &   125.26 &      ...  &    60.67 &      ...  &      ...  &      ...  &      ...  \\
  4267.826 &  Fe I &  3.11 & -1.17 &      ...  &      ...  &      ...  &      ...  &      ...  &      ...  &      ...  &      ...  &      ...  &      ...  &      ...  &      ...  &    80.59 &      ...  &      ...  &      ...  &      ...  \\
  4298.036 &  Fe I &  3.05 & -1.43 &      ...  &      ...  &      ...  &      ...  &      ...  &      ...  &      ...  &      ...  &      ...  &      ...  &      ...  &      ...  &    65.17 &      ...  &      ...  &      ...  &      ...  \\
  4352.734 &  Fe I &  2.22 & -1.29 &      ...  &      ...  &      ...  &      ...  &      ...  &      ...  &      ...  &      ...  &      ...  &      ...  &      ...  &      ...  &   105.35 &      ...  &      ...  &      ...  &      ...  \\
  4388.407 &  Fe I &  3.60 & -0.68 &      ...  &      ...  &      ...  &      ...  &   119.91 &      ...  &      ...  &      ...  &      ...  &      ...  &      ...  &      ...  &      ...  &      ...  &      ...  &      ...  &      ...  \\
  4422.568 &  Fe I &  2.85 & -1.11 &      ...  &      ...  &      ...  &      ...  &      ...  &      ...  &      ...  &      ...  &      ...  &      ...  &      ...  &      ...  &    82.99 &      ...  &      ...  &      ...  &      ...  \\
  4433.782 &  Fe I &  3.60 & -1.27 &      ...  &      ...  &      ...  &      ...  &    77.91 &      ...  &      ...  &      ...  &      ...  &      ...  &      ...  &      ...  &      ...  &      ...  &      ...  &      ...  &      ...  \\
  4442.339 &  Fe I &  2.20 & -1.25 &      ...  &      ...  &      ...  &      ...  &      ...  &      ...  &      ...  &      ...  &   147.93 &      ...  &      ...  &      ...  &   109.04 &      ...  &      ...  &      ...  &      ...  \\
  4443.194 &  Fe I &  2.86 & -1.04 &      ...  &   174.59 &      ...  &      ...  &      ...  &      ...  &      ...  &      ...  &      ...  &   137.16 &      ...  &      ...  &      ...  &      ...  &   142.63 &   135.23 &      ...  \\
  4446.832 &  Fe I &  3.69 & -1.32 &    84.17 &   102.26 &    95.20 &      ...  &      ...  &      ...  &      ...  &      ...  &      ...  &    87.80 &    98.86 &    87.51 &      ...  &      ...  &    93.48 &    76.39 &      ...  \\
  4447.717 &  Fe I &  2.22 & -1.34 &      ...  &      ...  &      ...  &      ...  &      ...  &      ...  &      ...  &      ...  &   139.32 &      ...  &      ...  &      ...  &   110.51 &      ...  &      ...  &      ...  &      ...  \\
  4476.018 &  Fe I &  2.85 & -0.82 &      ...  &   287.73 &      ...  &      ...  &      ...  &      ...  &      ...  &      ...  &      ...  &      ...  &      ...  &      ...  &      ...  &      ...  &      ...  &      ...  &      ...  \\
  4484.220 &  Fe I &  3.60 & -0.86 &      ...  &      ...  &      ...  &      ...  &      ...  &      ...  &      ...  &      ...  &      ...  &   102.09 &      ...  &   104.97 &    75.80 &   128.74 &   101.58 &      ...  &      ...  \\
  4595.358 &  Fe I &  3.30 & -1.76 &      ...  &      ...  &      ...  &      ...  &      ...  &      ...  &      ...  &      ...  &      ...  &      ...  &      ...  &      ...  &      ...  &      ...  &   121.51 &   101.64 &      ...  \\
  4643.463 &  Fe I &  3.65 & -1.15 &      ...  &      ...  &      ...  &      ...  &      ...  &      ...  &   119.68 &      ...  &    60.92 &    75.10 &      ...  &      ...  &    46.38 &      ...  &      ...  &      ...  &      ...  \\
  4647.434 &  Fe I &  2.95 & -1.35 &      ...  &      ...  &      ...  &      ...  &      ...  &      ...  &      ...  &      ...  &      ...  &   120.47 &      ...  &      ...  &    70.01 &      ...  &      ...  &      ...  &      ...  \\
  4736.773 &  Fe I &  3.21 & -0.75 &      ...  &      ...  &      ...  &      ...  &      ...  &      ...  &      ...  &      ...  &   143.63 &      ...  &      ...  &      ...  &    90.94 &      ...  &      ...  &      ...  &      ...  \\
  4786.807 &  Fe I &  3.02 & -1.61 &      ...  &      ...  &      ...  &      ...  &      ...  &      ...  &      ...  &      ...  &      ...  &      ...  &   116.44 &      ...  &      ...  &      ...  &      ...  &      ...  &      ...  \\
  4789.651 &  Fe I &  3.55 & -0.96 &   107.97 &   141.88 &      ...  &      ...  &      ...  &      ...  &      ...  &      ...  &      ...  &    99.93 &   112.76 &      ...  &      ...  &      ...  &      ...  &      ...  &      ...  \\
  4800.649 &  Fe I &  4.14 & -1.03 &      ...  &   108.89 &   113.31 &      ...  &    91.18 &   102.25 &    76.34 &    99.01 &      ...  &      ...  &    96.17 &    82.78 &      ...  &   106.41 &    93.72 &    79.11 &      ...  \\
  4872.137 &  Fe I &  2.88 & -0.57 &      ...  &      ...  &      ...  &      ...  &      ...  &      ...  &      ...  &      ...  &      ...  &      ...  &      ...  &      ...  &   121.11 &      ...  &      ...  &      ...  &      ...  \\
  4878.211 &  Fe I &  2.89 & -0.89 &      ...  &   411.41 &      ...  &      ...  &      ...  &      ...  &      ...  &      ...  &      ...  &      ...  &      ...  &      ...  &      ...  &      ...  &      ...  &      ...  &      ...  \\
  4903.310 &  Fe I &  2.88 & -0.93 &      ...  &   262.36 &      ...  &      ...  &      ...  &      ...  &      ...  &      ...  &      ...  &      ...  &      ...  &      ...  &   100.87 &      ...  &      ...  &      ...  &      ...  \\
  4930.315 &  Fe I &  3.96 & -1.20 &   108.30 &      ...  &      ...  &      ...  &      ...  &   128.39 &    97.20 &      ...  &    73.71 &    92.86 &   123.25 &   103.57 &    42.08 &   139.86 &   106.15 &   104.74 &      ...  \\
  4978.603 &  Fe I &  3.98 & -0.88 &   146.89 &      ...  &      ...  &      ...  &      ...  &      ...  &      ...  &      ...  &      ...  &      ...  &      ...  &      ...  &    69.07 &      ...  &      ...  &   129.47 &      ...  \\
  4985.253 &  Fe I &  3.93 & -0.56 &      ...  &      ...  &      ...  &      ...  &      ...  &      ...  &      ...  &      ...  &      ...  &      ...  &      ...  &      ...  &      ...  &      ...  &   120.35 &      ...  &      ...  \\
  5028.126 &  Fe I &  3.57 & -1.12 &      ...  &      ...  &      ...  &      ...  &      ...  &      ...  &      ...  &      ...  &      ...  &    82.09 &      ...  &      ...  &      ...  &      ...  &      ...  &      ...  &      ...  \\
  5049.819 &  Fe I &  2.28 & -1.35 &      ...  &      ...  &      ...  &      ...  &      ...  &      ...  &      ...  &      ...  &   120.87 &      ...  &      ...  &      ...  &   104.42 &      ...  &      ...  &      ...  &      ...  \\
  5068.766 &  Fe I &  2.94 & -1.04 &      ...  &      ...  &      ...  &      ...  &      ...  &      ...  &      ...  &      ...  &      ...  &      ...  &      ...  &      ...  &    93.18 &      ...  &      ...  &   138.99 &      ...  \\
  5171.596 &  Fe I &  1.48 & -1.79 &      ...  &      ...  &      ...  &      ...  &      ...  &      ...  &      ...  &      ...  &      ...  &      ...  &      ...  &      ...  &   113.11 &      ...  &      ...  &      ...  &      ...  \\
  5191.454 &  Fe I &  3.04 & -0.55 &      ...  &      ...  &      ...  &      ...  &      ...  &      ...  &      ...  &      ...  &      ...  &      ...  &      ...  &      ...  &   130.32 &      ...  &      ...  &      ...  &      ...  \\
  5202.336 &  Fe I &  2.18 & -1.84 &      ...  &   268.78 &      ...  &      ...  &      ...  &      ...  &      ...  &      ...  &      ...  &      ...  &      ...  &      ...  &      ...  &      ...  &      ...  &      ...  &      ...  \\
  5215.180 &  Fe I &  3.27 & -0.87 &      ...  &      ...  &      ...  &      ...  &      ...  &      ...  &      ...  &      ...  &      ...  &      ...  &      ...  &      ...  &    80.92 &      ...  &   127.05 &      ...  &      ...  \\
  5217.389 &  Fe I &  3.21 & -1.16 &      ...  &   160.38 &      ...  &      ...  &   139.81 &      ...  &   106.45 &      ...  &    92.95 &   116.85 &      ...  &      ...  &    67.99 &      ...  &   134.19 &   133.01 &      ...  \\
  5232.940 &  Fe I &  2.94 & -0.06 &      ...  &   583.58 &      ...  &      ...  &      ...  &      ...  &      ...  &      ...  &      ...  &      ...  &      ...  &      ...  &      ...  &      ...  &      ...  &      ...  &      ...  \\
  5242.490 &  Fe I &  3.63 & -0.97 &   114.26 &   116.82 &      ...  &    95.67 &   110.63 &   101.69 &    84.63 &      ...  &    75.05 &    84.52 &   112.38 &      ...  &    67.07 &   111.34 &    95.21 &    92.98 &    98.68 \\
  5253.462 &  Fe I &  3.28 & -1.57 &      ...  &      ...  &      ...  &      ...  &      ...  &      ...  &      ...  &      ...  &    62.78 &    80.26 &      ...  &      ...  &    40.72 &   106.22 &    89.83 &    84.18 &    56.66 \\
  5263.306 &  Fe I &  3.27 & -0.88 &      ...  &      ...  &      ...  &      ...  &      ...  &      ...  &   105.45 &      ...  &    97.24 &   127.22 &      ...  &      ...  &      ...  &      ...  &   138.26 &      ...  &      ...  \\
  5266.554 &  Fe I &  3.00 & -0.39 &      ...  &      ...  &      ...  &      ...  &      ...  &      ...  &      ...  &      ...  &      ...  &      ...  &      ...  &      ...  &   114.16 &      ...  &      ...  &      ...  &      ...  \\
  5281.789 &  Fe I &  3.04 & -0.83 &      ...  &      ...  &      ...  &      ...  &      ...  &      ...  &      ...  &      ...  &   120.32 &      ...  &      ...  &      ...  &    96.65 &      ...  &      ...  &      ...  &      ...  \\
  5283.621 &  Fe I &  3.24 & -0.53 &      ...  &      ...  &      ...  &      ...  &      ...  &      ...  &      ...  &      ...  &      ...  &      ...  &      ...  &      ...  &      ...  &      ...  &      ...  &      ...  &   144.75 \\
  5288.526 &  Fe I &  3.69 & -1.51 &    64.01 &    88.74 &    73.17 &    65.98 &    72.50 &    80.39 &    69.63 &   100.40 &    53.48 &    58.09 &    66.95 &      ...  &    30.98 &      ...  &    74.44 &    60.21 &    52.61 \\
  5302.300 &  Fe I &  3.28 & -0.72 &      ...  &   286.52 &      ...  &      ...  &      ...  &      ...  &      ...  &      ...  &   138.90 &      ...  &      ...  &      ...  &   116.98 &      ...  &      ...  &      ...  &   127.60 \\
  5324.178 &  Fe I &  3.21 & -0.10 &      ...  &      ...  &      ...  &      ...  &      ...  &      ...  &      ...  &      ...  &      ...  &      ...  &      ...  &      ...  &   142.67 &      ...  &      ...  &      ...  &      ...  \\
  5341.024 &  Fe I &  1.61 & -1.95 &      ...  &      ...  &      ...  &      ...  &      ...  &      ...  &      ...  &      ...  &      ...  &      ...  &      ...  &      ...  &   116.94 &      ...  &      ...  &      ...  &      ...  \\
  5365.399 &  Fe I &  3.57 & -1.02 &      ...  &      ...  &      ...  &    87.43 &      ...  &      ...  &      ...  &      ...  &      ...  &      ...  &      ...  &    90.08 &      ...  &      ...  &    87.52 &      ...  &    73.48 \\
  5367.465 &  Fe I &  4.42 &  0.44 &      ...  &   240.96 &      ...  &      ...  &      ...  &      ...  &   140.92 &      ...  &   137.66 &      ...  &      ...  &      ...  &    99.90 &      ...  &      ...  &      ...  &   127.12 \\
  5369.961 &  Fe I &  4.37 &  0.54 &      ...  &      ...  &      ...  &      ...  &      ...  &      ...  &      ...  &      ...  &      ...  &      ...  &      ...  &      ...  &   143.23 &      ...  &      ...  &      ...  &      ...  \\
  5379.573 &  Fe I &  3.69 & -1.51 &    74.69 &    92.71 &    71.24 &    65.15 &    75.48 &    82.60 &    68.03 &    79.51 &    52.64 &      ...  &    85.13 &      ...  &    34.58 &    84.68 &    80.91 &    72.30 &    43.29 \\
  5383.368 &  Fe I &  4.31 &  0.65 &      ...  &   33 ...  &      ...  &      ...  &      ...  &      ...  &      ...  &      ...  &      ...  &      ...  &      ...  &      ...  &   119.59 &      ...  &      ...  &      ...  &      ...  \\
  5393.167 &  Fe I &  3.24 & -0.71 &      ...  &      ...  &      ...  &      ...  &      ...  &      ...  &   137.60 &      ...  &      ...  &      ...  &      ...  &      ...  &   113.92 &      ...  &      ...  &      ...  &      ...  \\
  5410.909 &  Fe I &  4.47 &  0.40 &      ...  &      ...  &      ...  &      ...  &      ...  &      ...  &      ...  &      ...  &      ...  &      ...  &      ...  &      ...  &   122.89 &      ...  &      ...  &      ...  &   126.52 \\
  5412.783 &  Fe I &  4.43 & -1.72 &    37.81 &    33.61 &    38.86 &      ...  &    24.55 &      ...  &      ...  &      ...  &      ...  &    13.43 &    30.22 &      ...  &      ...  &    34.37 &    23.57 &    19.18 &     9.02 \\
  5415.198 &  Fe I &  4.39 &  0.64 &      ...  &   332.87 &      ...  &      ...  &      ...  &      ...  &      ...  &      ...  &      ...  &      ...  &      ...  &      ...  &   126.75 &      ...  &      ...  &      ...  &      ...  \\
  5429.696 &  Fe I &  0.96 & -1.88 &      ...  &   640.99 &      ...  &      ...  &      ...  &      ...  &      ...  &      ...  &      ...  &      ...  &      ...  &      ...  &      ...  &      ...  &      ...  &      ...  &      ...  \\
  5472.708 &  Fe I &  4.21 & -1.50 &    57.35 &    79.80 &    60.27 &    38.30 &    65.86 &    77.97 &    45.89 &    61.19 &      ...  &    43.15 &    76.56 &      ...  &    14.91 &    82.82 &    66.47 &    48.76 &    30.47 \\
  5525.543 &  Fe I &  4.23 & -1.08 &    69.94 &    83.49 &    94.70 &      ...  &    72.44 &    82.70 &      ...  &    75.63 &      ...  &    50.97 &    75.43 &    68.38 &      ...  &    88.21 &    64.35 &    53.78 &    40.55 \\
  5569.618 &  Fe I &  3.42 & -0.49 &      ...  &   360.87 &      ...  &      ...  &      ...  &      ...  &      ...  &      ...  &   136.34 &      ...  &      ...  &      ...  &    99.63 &      ...  &      ...  &      ...  &   117.47 \\
  5572.842 &  Fe I &  3.40 & -0.28 &      ...  &   500.82 &      ...  &      ...  &      ...  &      ...  &      ...  &      ...  &      ...  &      ...  &      ...  &      ...  &      ...  &      ...  &      ...  &      ...  &      ...  \\
  5586.755 &  Fe I &  3.37 & -0.14 &      ...  &      ...  &      ...  &      ...  &      ...  &      ...  &      ...  &      ...  &      ...  &      ...  &      ...  &      ...  &   119.36 &      ...  &      ...  &      ...  &      ...  \\
  5600.224 &  Fe I &  4.26 & -1.42 &      ...  &      ...  &      ...  &      ...  &      ...  &      ...  &      ...  &      ...  &      ...  &      ...  &      ...  &      ...  &    24.90 &      ...  &      ...  &      ...  &      ...  \\
  5624.542 &  Fe I &  3.42 & -0.76 &      ...  &      ...  &      ...  &   135.41 &      ...  &      ...  &      ...  &      ...  &      ...  &      ...  &      ...  &      ...  &      ...  &      ...  &      ...  &      ...  &   125.98 \\
  5661.345 &  Fe I &  4.28 & -1.76 &    26.65 &    50.18 &    31.82 &    20.27 &    34.85 &    42.75 &    26.46 &      ...  &      ...  &    19.04 &    43.79 &      ...  &      ...  &    41.86 &      ...  &    21.21 &    10.85 \\
  5662.516 &  Fe I &  4.18 & -0.57 &      ...  &      ...  &      ...  &      ...  &      ...  &      ...  &      ...  &      ...  &      ...  &      ...  &      ...  &      ...  &      ...  &      ...  &      ...  &      ...  &    74.70 \\
  5686.530 &  Fe I &  4.55 & -0.45 &   125.11 &      ...  &      ...  &      ...  &      ...  &      ...  &      ...  &      ...  &      ...  &      ...  &      ...  &      ...  &    62.22 &      ...  &      ...  &      ...  &    71.94 \\
  5705.464 &  Fe I &  4.30 & -1.35 &    49.65 &    59.30 &    41.62 &      ...  &    50.24 &    58.22 &    43.15 &    51.37 &    29.85 &    38.13 &    60.19 &    69.53 &      ...  &    61.17 &    53.21 &    43.81 &    24.61 \\
  5753.122 &  Fe I &  4.26 & -0.69 &      ...  &      ...  &      ...  &    78.88 &    81.56 &      ...  &   100.54 &      ...  &    65.01 &    82.85 &      ...  &    92.39 &      ...  &   107.86 &    88.82 &    79.55 &    70.17 \\
  5816.373 &  Fe I &  4.55 & -0.60 &   100.44 &      ...  &    92.40 &      ...  &    93.62 &      ...  &      ...  &      ...  &    72.18 &    84.85 &      ...  &      ...  &    41.81 &   131.62 &      ...  &   103.34 &    64.59 \\
  5855.075 &  Fe I &  4.61 & -1.48 &    31.35 &    49.11 &    32.43 &    25.70 &    36.54 &    40.94 &    31.70 &    34.63 &    22.02 &    26.81 &    41.16 &      ...  &      ...  &    38.61 &    44.34 &    33.57 &    17.99 \\
  6065.481 &  Fe I &  2.61 & -1.53 &      ...  &   198.39 &      ...  &   127.52 &      ...  &      ...  &   127.96 &      ...  &      ...  &   120.98 &      ...  &   137.15 &    89.41 &      ...  &   146.82 &      ...  &   103.09 \\
  6136.615 &  Fe I &  2.45 & -1.40 &      ...  &      ...  &      ...  &      ...  &      ...  &      ...  &      ...  &      ...  &      ...  &      ...  &      ...  &      ...  &   105.32 &      ...  &      ...  &      ...  &      ...  \\
  6137.691 &  Fe I &  2.59 & -1.40 &      ...  &      ...  &      ...  &      ...  &      ...  &      ...  &   128.34 &      ...  &   120.27 &      ...  &      ...  &      ...  &      ...  &      ...  &      ...  &      ...  &   124.09 \\
  6141.730 &  Fe I &  3.60 & -1.46 &      ...  &   140.70 &      ...  &      ...  &      ...  &   123.62 &      ...  &      ...  &   127.44 &      ...  &   132.14 &   138.04 &      ...  &      ...  &      ...  &   134.38 &      ...  \\
  6230.722 &  Fe I &  2.56 & -1.28 &      ...  &      ...  &      ...  &   161.20 &      ...  &      ...  &      ...  &      ...  &   126.16 &      ...  &      ...  &      ...  &   102.56 &      ...  &      ...  &      ...  &   142.28 \\
  6232.640 &  Fe I &  3.65 & -1.22 &      ...  &      ...  &      ...  &    85.66 &   104.07 &   117.82 &    88.26 &      ...  &    77.17 &    92.98 &   126.77 &    98.08 &    50.45 &   129.62 &   117.10 &   103.07 &    72.33 \\
  6246.318 &  Fe I &  3.60 & -0.88 &      ...  &      ...  &      ...  &   104.33 &      ...  &      ...  &   122.26 &      ...  &   110.72 &      ...  &      ...  &      ...  &    69.51 &      ...  &   137.91 &      ...  &   109.99 \\
  6252.555 &  Fe I &  2.40 & -1.69 &      ...  &      ...  &      ...  &   126.73 &      ...  &      ...  &   123.51 &      ...  &   111.30 &   129.45 &      ...  &   144.20 &    90.17 &      ...  &      ...  &   140.35 &   114.03 \\
  6301.500 &  Fe I &  3.65 & -0.72 &      ...  &   222.93 &      ...  &   118.53 &      ...  &      ...  &   120.32 &      ...  &      ...  &   121.26 &      ...  &   145.23 &    76.52 &      ...  &      ...  &      ...  &   108.21 \\
  6336.823 &  Fe I &  3.69 & -0.86 &      ...  &   183.21 &      ...  &   112.33 &      ...  &      ...  &   108.46 &      ...  &    98.75 &   117.06 &      ...  &   123.89 &    68.33 &      ...  &   132.60 &      ...  &    97.65 \\
  6400.000 &  Fe I &  3.60 & -0.29 &      ...  &      ...  &      ...  &      ...  &      ...  &      ...  &      ...  &      ...  &      ...  &      ...  &      ...  &      ...  &   126.24 &      ...  &      ...  &      ...  &      ...  \\
  6408.017 &  Fe I &  3.69 & -1.02 &      ...  &      ...  &      ...  &    88.44 &      ...  &      ...  &    92.80 &      ...  &      ...  &   125.52 &      ...  &   120.48 &      ...  &      ...  &      ...  &      ...  &    84.68 \\
  6411.648 &  Fe I &  3.65 & -0.72 &      ...  &      ...  &      ...  &   126.13 &      ...  &      ...  &   123.36 &      ...  &      ...  &      ...  &      ...  &      ...  &    81.65 &      ...  &      ...  &      ...  &   121.48 \\
  6494.980 &  Fe I &  2.40 & -1.27 &      ...  &      ...  &      ...  &      ...  &      ...  &      ...  &      ...  &      ...  &      ...  &      ...  &      ...  &      ...  &   104.32 &      ...  &      ...  &      ...  &   115.45 \\
  6677.985 &  Fe I &  2.69 & -1.42 &      ...  &      ...  &      ...  &   133.98 &      ...  &      ...  &   127.67 &      ...  &   115.30 &      ...  &      ...  &      ...  &    99.22 &      ...  &      ...  &      ...  &   119.38 \\
  6752.707 &  Fe I &  4.64 & -1.20 &    57.27 &    78.80 &      ...  &    4 ...  &    36.84 &    61.11 &    44.83 &      ...  &    28.82 &    41.43 &    73.59 &    42.40 &      ...  &    80.97 &    64.29 &    34.32 &    25.40 \\
  6803.999 &  Fe I &  4.65 & -1.50 &      ...  &      ...  &      ...  &      ...  &      ...  &      ...  &      ...  &      ...  &      ...  &      ...  &      ...  &    25.05 &      ...  &      ...  &      ...  &      ...  &      ...  \\
  6804.270 &  Fe I &  4.58 & -1.81 &      ...  &      ...  &      ...  &      ...  &      ...  &      ...  &      ...  &      ...  &      ...  &      ...  &    28.89 &      ...  &      ...  &      ...  &      ...  &      ...  &      ...  \\
  6837.005 &  Fe I &  4.59 & -1.69 &    18.19 &      ...  &      ...  &    22.11 &    21.74 &    26.49 &    22.19 &    19.13 &    13.73 &      ...  &    29.86 &    20.47 &      ...  &    27.54 &    23.93 &    13.99 &      ...  \\
  6854.823 &  Fe I &  4.59 & -1.93 &      ...  &      ...  &      ...  &      ...  &      ...  &      ...  &      ...  &      ...  &      ...  &      ...  &    20.70 &      ...  &      ...  &      ...  &      ...  &      ...  &      ...  \\
  8327.055 &  Fe I &  2.20 & -1.52 &      ...  &      ...  &      ...  &      ...  &      ...  &      ...  &      ...  &      ...  &   158.13 &      ...  &      ...  &      ...  &   119.23 &      ...  &      ...  &      ...  &   144.69 \\
  8387.771 &  Fe I &  2.18 & -1.49 &      ...  &   351.23 &      ...  &      ...  &      ...  &      ...  &      ...  &      ...  &      ...  &      ...  &      ...  &      ...  &   117.33 &      ...  &      ...  &      ...  &      ...  \\
  8598.828 &  Fe I &  4.39 & -1.09 &    64.80 &    81.51 &      ...  &    53.08 &    66.11 &      ...  &    60.62 &    73.10 &    45.87 &      ...  &      ...  &      ...  &      ...  &      ...  &    75.29 &    72.86 &    55.77 \\
  8688.623 &  Fe I &  2.18 & -1.21 &      ...  &      ...  &      ...  &      ...  &      ...  &      ...  &      ...  &      ...  &      ...  &      ...  &      ...  &      ...  &   143.88 &      ...  &      ...  &      ...  &      ...  \\
  4178.854 & Fe II &  2.58 & -2.44 &      ...  &      ...  &    53.64 &      ...  &      ...  &      ...  &      ...  &      ...  &      ...  &    88.52 &      ...  &    81.38 &   114.92 &    88.57 &   125.02 &      ...  &      ...  \\
  4491.397 & Fe II &  2.86 & -2.64 &      ...  &      ...  &      ...  &      ...  &      ...  &      ...  &   112.76 &      ...  &      ...  &      ...  &      ...  &    60.18 &   100.49 &    71.21 &    76.85 &    86.18 &      ...  \\
  4515.333 & Fe II &  2.84 & -2.36 &      ...  &      ...  &      ...  &      ...  &      ...  &      ...  &   126.08 &      ...  &   120.31 &   115.81 &      ...  &    89.17 &      ...  &      ...  &   134.19 &   108.13 &      ...  \\
  4555.887 & Fe II &  2.83 & -2.25 &      ...  &      ...  &      ...  &      ...  &      ...  &      ...  &      ...  &      ...  &      ...  &      ...  &      ...  &      ...  &      ...  &    75.34 &      ...  &      ...  &      ...  \\
  4576.333 & Fe II &  2.84 & -2.92 &      ...  &    76.85 &    28.28 &      ...  &    59.23 &      ...  &    86.72 &      ...  &    78.47 &    59.65 &    73.84 &    45.71 &      ...  &    70.87 &    89.25 &      ...  &      ...  \\
  4583.829 & Fe II &  2.81 & -1.74 &      ...  &   190.79 &   138.57 &      ...  &      ...  &      ...  &      ...  &      ...  &      ...  &      ...  &      ...  &      ...  &      ...  &   132.52 &      ...  &      ...  &      ...  \\
  4629.331 & Fe II &  2.81 & -2.26 &   103.92 &   125.22 &      ...  &      ...  &   117.65 &      ...  &   128.34 &   109.82 &   103.78 &   108.27 &   120.93 &   101.77 &   113.57 &   122.24 &   125.42 &   109.03 &      ...  \\
  5316.609 & Fe II &  3.15 & -1.78 &   126.06 &      ...  &    81.27 &      ...  &   132.53 &      ...  &      ...  &      ...  &      ...  &      ...  &      ...  &   138.39 &      ...  &      ...  &      ...  &      ...  &      ...  \\
  7711.720 & Fe II &  3.90 & -2.45 &    25.80 &      ...  &     8.22 &    60.08 &    19.27 &    56.95 &    77.48 &    27.63 &    58.79 &    45.31 &    40.88 &    33.48 &    66.91 &    47.99 &    57.19 &    42.85 &    47.16 \\
\enddata
\end{deluxetable}
\end{longrotatetable}
\restoregeometry

\newgeometry{bmargin=-1in}

\begin{longrotatetable}
\begin{deluxetable}{|c c c c c c c c c c c c c|}

\tablecaption{Central effective temperature, log(g), and [Fe/H] values derived by this work and by literature sources \citep{Petigura+Marcy2011,BreweCO,Sousa+2021} The bottom row is the average difference between the reference and this work, with the error term being the standard deviation of the residuals. \label{table:Paramcomp}}
\tabletypesize{\scriptsize}
\tablenum{8}
\tablehead{
\hline
\colhead{} &
\multicolumn{3}{c}{This work} &
\multicolumn{3}{c}{P\&M 2011} &
\multicolumn{3}{c}{B\&F 2016} &
\multicolumn{3}{c}{Sousa et al. 2021} \\
\colhead{Star Name} &
\colhead{\teff} &
\colhead{log(g)} &
\colhead{[Fe/H]} &
\colhead{\teff} &
\colhead{log(g)} &
\colhead{[Fe/H]} &
\colhead{\teff} &
\colhead{log(g)} &
\colhead{[Fe/H]} &
\colhead{\teff} &
\colhead{log(g)} &
\colhead{[Fe/H]} 
}
\startdata
18 Eridani & 5097$\pm$50 & 4.58$\pm$0.02 & -0.05$\pm$0.10 & ... & ... & ... & 5065 & 4.55 & -0.01 & 5053$\pm$46 & 4.45$\pm$0.10 & -0.14$\pm$0.03\\
55 Cancri & 5308$\pm$40 & 4.46$\pm$0.03 & 0.30$\pm$0.06 &  ... & ... & ... & 5250 & 4.36 & 0.35 & 5363$\pm$59 & 4.28$\pm$0.14 & 0.33$\pm$0.04\\
HAT-P-1 & 5812$\pm$190 & 4.26$\pm$0.09 & 0.01$\pm$0.11 &  ... & ... & ... & 5964 & 4.32 & 0.16 & 6074$\pm$16 & 4.41$\pm$0.03 & 0.20$\pm$0.01\\
HAT-P-26 & 5289$\pm$80 & 4.52$\pm$0.02 & 0.02$\pm$0.10 &  ... & ... & ... & 5039 & 4.45 & 0.05 & 5001$\pm$61 & 4.21$\pm$0.15 & 0.01$\pm$0.04\\
HD 80606 & 5547$\pm$40 & 4.37$\pm$0.04 & 0.19$\pm$0.09 & 5573 & 4.44 & 0.26 & 5524 & 4.31 & 0.31 & 5581$\pm$29 & 4.33$\pm$0.07 & 0.33 $\pm$0.02\\    
HD 149026 & 6029$\pm$20 & 4.20$\pm$0.02 & 0.31$\pm$0.09 & ... & ... & ... & 6084 & 4.24 & 0.35 & 6166$\pm$32 & 4.35$\pm$0.05 & 0.37$\pm$0.02\\
HD 189733 & 5099$\pm$30 & 4.56$\pm$0.02 & -0.12$\pm$0.08 & ... & ... & ... & 5023 & 4.51 & 0.06 & 4969$\pm$48 & 4.30$\pm$0.12 & -0.08$\pm$0.03\\
HD 209458 & 6031$\pm$20 & 4.31$\pm$0.02 & -0.01$\pm$0.06 & 6099 & 4.37 & 0.02 & 6052 & 4.34 & 0.05 & 6126$\pm$18 & 4.50$\pm$0.04 & 0.04$\pm$0.01\\
Kepler-51 & 5577$\pm$40 & 4.46$\pm$0.02 & -0.29$\pm$0.07 & ... & ... & ... & ... & ... & ... & 5673$\pm$60* & 4.70$\pm$0.10* & 0.05$\pm$0.04* \\
TOI 421 & 5324$\pm$60 & 4.52$\pm$0.02 & -0.03$\pm$0.08 & ... & ... & ... & ... & ... & ... & 5316$\pm$27 & 4.35$\pm$0.06 & -0.03$\pm$0.02 \\
WASP-17 & 6157$\pm$150 & 4.02$\pm$0.05 & -0.30$\pm$0.11 & ... & ... & ... & ... & ... & ... & 6793$\pm$52 & 4.59$\pm$0.04 & -0.01$\pm$0.03 \\
WASP-52 & 5121$\pm$60 & 4.55$\pm$0.03 & 0.08$\pm$0.09 & ... & ... & ... & ... & ... & ... & 5073$\pm$63 & 4.36$\pm$0.15 & 0.17$\pm$0.03 \\
WASP-63 & 5512$\pm$80 & 3.94$\pm$0.06 & -0.01$\pm$0.15 & ... & ... & ... & ... & ... & ... & 5676$\pm$40 & 4.12$\pm$0.06 & 0.26$\pm$0.03 \\
WASP-77A & 5660$\pm$80 & 4.49$\pm$0.03 & -0.15$\pm$0.06 & ... & ... & ... & ... & ... & ... & 5595$\pm$16 & 4.41$\pm$0.03 & 0.01$\pm$0.01 \\
WASP-127 & 5949$\pm$60 & 4.24$\pm$0.02 & -0.35$\pm$0.05 & ... & ... & ... & ... & ... & ... & 5832$\pm$14 & 4.31$\pm$0.03 & -0.18$\pm$0.01 \\
\hline
\textbf{Avg. Deviation} & ... & ... & ... & \textbf{47$\pm$21} & \textbf{0.07$\pm$0.01} & \textbf{0.05$\pm$0.02} & \textbf{-26$\pm$108} & \textbf{-0.02$\pm$0.05} & \textbf{0.08$\pm$0.05} & \textbf{52$\pm$204} & \textbf{0.01$\pm$0.22} & \textbf{0.08$\pm$0.13}
\enddata
\tablecomments{We did not find any common stars between our sample and that of \citet{Nissen2013}}
\tablecomments{Kepler-51's data in SWEET-Cat was taken from \citet{Petigura+2017}}

\end{deluxetable}
\end{longrotatetable}

\end{CJK*}
\end{document}